\documentclass[onecolumn,prd,aps,floats,floatfix,nofootinbib,amsmath,amssymb]{revtex4}
\usepackage{graphicx}
\usepackage{color}
\usepackage{amsmath,amssymb}
\usepackage{array}

\newcommand{\vev}[1]{\langle #1 \rangle}

\newcommand{\TeV}{\text{\,TeV}}
\newcommand{\GeV}{\text{\,GeV}}

\newcommand{\gtwo}{I\kern-.1em I\,}
\newcommand{\be}{\begin{equation}}
\newcommand{\ee}{\end{equation}}
\newcommand{\beq}{\begin{eqnarray}}
\newcommand{\eeq}{\end{eqnarray}}
\newcommand{\bpm}{\begin{pmatrix}}
\newcommand{\epm}{\end{pmatrix}}
\newcommand{\cl}{\, \rm C.L.}

\begin{document}

\title{Top-seesaw assisted technicolor model and a $m=126$ GeV scalar}
\author{Hidenori S. Fukano}
\email{hidenori.f.sakuma@jyu.fi} 
\author{Kimmo Tuominen}
\email{kimmo.i.tuominen@jyu.fi}
\affiliation{Department of Physics, University of Jyv\"askyl\"a, P.O.Box 35, FIN-40014 Jyv\"askyl\"a, Finland \\
and 
Helsinki Institute of Physics, P.O.Box 64, FIN-00014 University of Helsinki, Finland\\}

\begin{abstract}
We consider a model of strong dynamics able to account for the origin of the electroweak symmetry
breaking and heavy quark masses. The model is based on a technicolor sector, augmented 
with topcolor and top-seesaw mechanism
to assist in the generation of heavy quark masses.
The low energy effective theory is a particular three Higgs doublet model. The additional feature
is the possibility of the existence of composite higher spin states beyond the scalars, which are 
shown to be essential  in this model to provide extra contributions in the higgs decays into two photons. 
We provide a detailed strategy and analysis how this type of models are to be constrained with the present data.
\end{abstract}

\maketitle

%
\section{Introduction}
%
The ATLAS \cite{:2012gk} and CMS \cite{:2012gu} experiments at LHC have announced a discovery of a new boson with mass $M_h\simeq 126$ GeV. The decay and production rates of this new particle appear to be consistent with the prediction of the Standard Model (SM) of
elementary particle interactions, and therefore the next logical step is to try to 
uncover its properties more precisely and to see how well it fits in with various
extensions of the Standard Model. For examples, see e.g. \cite{Ellis:2012hz,Espinosa:2012im,Espinosa:2012in,Matsuzaki:2012mk,Elander:2012fk,Frandsen:2012rj,Chivukula:2012cp,Low:2012rj,Coleppa:2012eh,Eichten:2012qb}.

Strong dynamics remains as a viable alternative, although the discovery of a light 
scalar particle is a severe obstruction for traditional Technicolor models \cite{Susskind:1978ms,Weinberg:1979bn}.
Moreover, technicolor alone does not provide a mechanism to generate 
masses for the elementary matter fermions, and one must invoke more complex
dynamical mechanisms. 

How does a light scalar emerge from strongly coupled dynamics ? There are generally at least two different alternatives. First possibility arises, if the theory underlying the dynamical electroweak symmetry breaking is quasiconformal \cite{Holdom:1984sk,Yamawaki:1985zg,Akiba:1985rr,Appelquist:1986an}. This means that under the renormalization group evolution the theory approaches an infrared fixed point which, however, is supercritical with respect to chiral symmetry breaking; formation of fermion-antifermion condensate triggers electroweak symmetry breaking and the theory flows into QCD like vacuum in the deep infrared. However, due to the presence of a quasi stable infrared fixed point the coupling constant evolves very slowly, i.e. walks, over a large hierarchy of scales and this quasiconformal behavior is directly reflected on the properties of the spectrum \cite{Sannino:2004qp, Dietrich:2005jn}. The second alternative is that the contributions to the electroweak sector are shared beteween different sectors, i.e. there exists different scales, say $v_1$ and $v_2$ which together give $v_{\textrm{weak}}^2=v_1^2+v_2$, but both $v_1$ and $v_2$ can be less than $v_{\textrm{weak}}$. The masses of the excitations in different sectors are dictated by the scales $v_1$ and $v_2$, and hence these mass scales can also be smaller than 
$v_{\textrm{weak}}$. This latter possibility will be considered in this paper.

Models of this type are motivated by the need to explain both the generation of the masses of the electroweak gauge bosons as well as the masses of the elementary fermion fields of the SM. 
We  assume that the light fermion masses are explained by some Extended Technicolor (ETC) scenario  \cite{Dimopoulos:1979es}, while the masses of the third generation quarks arise dominantly from additional strong dynamics, which we assume to be of top-seesaw type \cite{Dobrescu:1997nm,Chivukula:1998wd,He:2001fz} in \cite{Fukano:2011fp,Fukano:2012qx}.

The top-seesaw sector in this model is based on \cite{He:2001fz}.
The basic idea of this model building is to introduce new vectorlike quarks not
charged under the weak interaction, but which generate a nontrivial vacuum condensate via new strong dynamics shared between the third and fourth generation quarks. Concretely, under SU(3)$_1\times$SU(3)$_2\times$
SU$_L\times$U(1)$_1\times$U(1)$_2$, these fields transform as
\beq
Q_L^{(3)}& \sim& (3,1,2,1/6,0),\qquad U_R^{(3)}\sim (1,1,3,1,0,2/3),\quad 
U_R^{(3)}\sim (1,1,3,1,0,-1/3)\nonumber \\
U_L^{(4)}&\sim& (1,1,3,1,0,2/3),\quad U_R^{(4)}\sim (1,3,1,1,2/3,0),
\quad D_L^{(4)}\sim (1,1,3,1,0,-1/3),\quad D_R^{(4)}\sim
(1,3,1,1,-1/3,0).
\eeq
The full underlying gauge symmetry is assumed to reduce to 
SU(3)$_{\textrm{QCD}}\times$SU(2)$_L\times$U(1)$_Y$ via symmetry breaking at scale $\Lambda\gg v_{\textrm{weak}}$, generating  effective four fermion interactions 
\beq
{\cal L}^{4f}
=
G_b \left( \bar{D}^{(4)}_R Q^{(3)}_L\right)^2 
+
G_t \left( \bar{U}^{(4)}_R Q^{(3)}_L\right)^2 
+
G_{tb} \left( \bar{Q}^{(3)}_L U^{(4)}_R \right) 
\left( \bar{D}^{(4) c}_R i \tau_2 Q^{(3) c}_L\right)
\,,
\label{4f-present}
\eeq
where the superscript $^c$ implies charge conjugation. 
The diagonal terms, $G_b$ and $G_t$  arise from the exchange of eight colored
gauge bosons 
with mass $\sim\Lambda$. The off diagonal term $G_{tb}$ may arise either from FCNC interactions
of the topcolor \cite{Hill:1991at} sector or via the topcolor instantons \cite{He:2001fz}.

The above four-fermion interactions, when strong, lead to vacuum condensates of the form
$\bar{D}_R^{(4)}Q_L^{(3)}$ and $\bar{U}_R^{(4)}Q_L^{(3)}$, and these contribute to 
electroweak symmetry breaking and to the masses of heavy quarks. The quark mass spectrum is fully specified by noting that in addition to the dynamically generated 
quark masses $\Sigma_U$ and $\Sigma_D$, the low energy gauge invariance also
allows for the mass terms $M_U^{(43)}\bar{U}_L^{(4)}U_R^{(3)}$ and $M_U^{(44)}\bar{U}_L^{(4)}U_R^{(4)}$ and similarly with the replacement $U\rightarrow D$. 
The quark mass spectrum is then determined from
\be
{\cal{L}}_M=-(\bar{U}_L^{(3)},\bar{U}_L^{(4)})\left(\begin{array}{cc} 
0 & \Sigma_{U} \\ M_U^{(43)} & M_U^{(44)}\end{array}\right)\left(\begin{array}{c}
U_R^{(3)} \\ U_R^{(4)}\end{array}\right)
-(\bar{D}_L^{(3)},\bar{D}_L^{(4)})\left(\begin{array}{cc} 
0 & \Sigma_{D} \\ M_D^{(43)} & M_D^{(44)}\end{array}\right)\left(\begin{array}{c}
D_R^{(3)} \\ D_R^{(4)}\end{array}\right)+{\textrm{h.c.}}.
\ee

From the structure of the condensates, it follows that the top-seesaw sector is, at
low energies, described by an effective two-higgs doublet model. 
As in \cite{Fukano:2012qx}, we consider also an underlying (extended) technicolor sector responsible for 
the electroweak  symmetry breaking, but contributing only in subleading order to the heavy quark masses. 
The light fermion masses are expected to be generated by the extended technicolor interactions. 
Considering technicolor and top-seesaw dynamics together then 
leads to a three-doublet model as an effective low energy description of the strong dynamics.

In \cite{Fukano:2012qx} we considered a concrete model built upon the minimal
walking technicolor \cite{Sannino:2004qp, Dietrich:2005jn} model. Here, we keep the 
technicolor sector generic, with the chiral symmetry SU(2)$_L\times$
SU(2)$_R$ chiral symmetry whose spontaneous breaking contributes to electroweak symmetry breaking. 
When comparing with precision data, we also
outline  how different technicolor models, like the minimal walking technicolor, affect the results.

The paper is organized as follows: First we introduce the low energy Lagrangian 
in section \ref{sec:effLagr} . Then, in section \ref{dynamical-results} we 
compute the spectrum of fermions and composite particles. The constraints from 
electroweak precision observables are considered in section \ref{EWPT-gLb}, and finally 
in section \ref{126higgsLHC} we consider the model in light of the recent LHC data.

%
\section{Low energy effective Lagrangian}
\label{sec:effLagr}
%

%
In this section, we consider the low energy effective theory for the top-seesaw assisted TC model. To describe the Nambu-Goldstone bosons
(NGBs) of the TC sector, we use the most minimal electroweak chiral Lagrangian 
(EWCL) 
\cite{Appelquist:1980vg,Longhitano:1980iz,Longhitano:1980tm} based on  the 
$G/H = [SU(2)_L \times SU(2)_R]/SU(2)_V$, 
which is the most minimal structure. In other words, 
the leading order chiral Lagrangian is
\beq
{\cal L}^{(2)}_{\rm EWCL}
=
\left| D_\mu \Phi_{\rm TC} \right|^2 
\,,
\label{chiral-Lag-kin}
\eeq
where $\Phi_{\rm TC}$ is  given by
\beq
\Phi_{\rm TC} = \bpm \pi^+_{\rm TC} \\[2ex] \dfrac{1}{\sqrt{2}}\left[ v_{\rm TC} - i \pi^0_{\rm TC}\right] \epm\, ,
\label{TC-higgs}
\eeq
and $\tilde{\Phi}_{\rm TC} \equiv i \tau^2 \Phi^*_{\rm TC}$, where $\tau^2$ is the second Pauli matrix.
The covariant derivative $D_\mu \Phi_{\rm TC}$ is given by
\beq
D_\mu \Phi_{\rm TC}
=
\partial_\mu \Phi_{\rm TC} -  i  g W^a_\mu T^a \Phi_{\rm TC} - \frac{1}{2} g' B_\mu \Phi_{\rm TC}\,,
\label{higgs-covariant-derivative}
\eeq
where $T^a = (1/2) \tau^a$ and  $W_\mu,B_\mu$ are the SM $SU(2)_L,U(1)_Y$ gauge boson fields and $g,g'$ are their gauge couplings. 
On the other hand, the top-seesaw sector is described by the two higgs doublet model (2HDM) \cite{He:2001fz}, 
i.e. by doublets $\Phi_i\,(i=1,2)$
\beq
\Phi_i = \bpm \pi^+_i \\[1ex] \dfrac{1}{\sqrt{2}} \left[ v_i + h^0_i - i \pi^0_i \right]\epm\,,
\label{TSS-higgs}
\eeq
and the covariant derivatives for $\Phi_i$ under the electroweak gauge symmetry are as in Eq.(\ref{higgs-covariant-derivative}). Thus the low energy effective Lagrangian of the top-seesaw assisted TC model is given by
\beq
{\cal L}_{\rm higgs} (\Phi_1,\Phi_2,\Phi_{\rm TC})
=
\sum_{i=1,2,{\rm TC}} \left| D_\mu \Phi_i \right|^2 
+ {\cal L}_{\rm yukawa}
- V(\Phi_1,\Phi_2,\Phi_{\rm TC}) \,.
\label{hybrid-full-EFT}
\eeq
Here, ${\cal L}_{\rm yukawa}$ consists of the Yukawa interaction terms and is given explicitly by
\beq
{\cal L}_{\rm yukawa}
\!\!&=&\!\!
- \!\!\!\sum^{\text{quarks}}_{i,j=1,2,3} y^{(d)}_{ij} \bar{Q}^{(i)}_L \Phi_{\rm TC} D^{(j)}_R
- \!\!\!\sum^{\text{quarks}}_{i,j=1,2,3} y^{(u)}_{ij} \bar{Q}^{(i)}_L \tilde{\Phi}_{\rm TC} U^{(j)}_R
\nonumber\\[1ex]
&&
-  y_1 \bar{Q}^{(3)}_L \Phi_1 D^{(4)}_R 
- y_2 \bar{Q}^{(3)}_L \tilde{\Phi}_2 U^{(4)}_R 
\nonumber\\[1ex]
&&
+ {\rm h.c. }
\,.
\label{reno-yukawa-H4G}
\eeq
Note here that the implications from the Yukawa term in Eq.(\ref{reno-yukawa-H4G}) 
are very different from the usual 
2HDM \cite{Branco:2011iw}.  This is so since the neutral higgs boson $h^0$ arises 
only from the doublets $\Phi_{1,2}$ 
of Eqs.(\ref{TC-higgs}) and (\ref{TSS-higgs}), does not couple to any leptons or any 
light quarks at tree level. 
Therefore, for the phenomenological purposes we will concentrate only on the quark 
sector and we will omit the light generations
in what follows. Also note that due to this underlying structure, the FCNC problem of 
the generic 2HDM is completely avoided in our model.
%
%

The potential $V(\Phi_1,\Phi_2,\Phi_{\rm TC})$ in Eq.(\ref{hybrid-full-EFT}) arising from the top-seesaw sector can be decomposed as 
\beq
V(\Phi_1,\Phi_2,\Phi_{\rm TC}) 
=
V_{\rm TSS}(\Phi_1,\Phi_2) 
+
V_M(\Phi_1,\Phi_2,\Phi_{\rm TC})
\label{3HDM-potential}
\,.
\eeq
We take $V_{\rm TSS}(\Phi_1,\Phi_2)$ to be of the form
\beq
V_{\rm TSS}(\Phi_1,\Phi_2) 
&=& 
M^2_{11} |\Phi_1|^2 + M^2_{22} |\Phi_2|^2 
- M^2_{12} \left[ \Phi^\dagger_1 \Phi_2 + {\rm h.c.}\right] 
\nonumber\\[1ex]
&&
+ \frac{1}{2} \lambda_1( \Phi^\dagger_1 \Phi_1 )^2 
+ \frac{1}{2} \lambda_2 ( \Phi^\dagger_2 \Phi_2)^2 
+ \lambda_3(\Phi^\dagger_1 \Phi_1)(\Phi^\dagger_2 \Phi_2)
+ \lambda_4(\Phi^\dagger_1 \Phi_2)(\Phi^\dagger_2 \Phi_1)
\,.
\label{TSS-higgs-potential}
\eeq
All these terms are generated by the underlying theory via the four fermion interactions (\ref{4f-present}).

Note that this scalar potential for top-seesaw sector in Eq.(\ref{TSS-higgs-potential}) 
is different from the scalar potential given in \cite{Fukano:2012qx,He:2001fz}. In \cite{Fukano:2012qx,He:2001fz}, there are $[\lambda_6 ( \Phi^\dagger_1 \Phi_1 ) + \lambda_7 ( \Phi^\dagger_2 \Phi_2)] (\Phi^\dagger_1 \Phi_2)+ {\rm h.c.} ]$-terms, 
which arise from the Peccei-Quinn (PQ) $U(1)_A$ symmetry is breaking topcolor 
instanton induced four fermion interaction. Here, on the other hand, we do not 
specify the PQ-symmetry breaking mechanism, but  assume instead that PQ-
symmetry breaks by $M^2_{12}$-term  derived from the last term in 
Eq.(\ref{4f-present}) \cite{Miransky:1988xi,Luty:1990bg,Hashimoto:2009ty}. 
In comparison to generic two-doublet models we remark, 
that the potential Eq.(\ref{TSS-higgs-potential}) is derived by the bubble-sum approximation \cite{Bardeen:1989ds} from the microscopic Lagrangian (\ref{4f-present}), and hence does not include $\lambda_5 [( \Phi^\dagger_1 \Phi_2 )^2 + \text{h.c.}]$ term \cite{Hashimoto:2009ty,Fukano:2012qx}.
%
%
%

To account for the mixing between the TC sector and the top-seesaw sector, we have added  $V_M(\Phi_1,\Phi_2,\Phi_{\rm TC})$ to the 2HDM potential Eq.(\ref{TSS-higgs-potential}) in Eq.(\ref{3HDM-potential}). This contribution is given by \cite{Chivukula:2011ag}
\beq
V_M(\Phi_1,\Phi_2,\Phi_{\rm TC})
\!\!\!&=&\!\!\!
c_1 v^2_1  \left| \Phi_1- \frac{v_1}{v_{\rm TC}}\Phi_{\rm TC} \right|^2 
+ c_2 v^2_2 \left| \Phi_2- \frac{v_2}{v_{\rm TC}}\Phi_{\rm TC} \right|^2\,,
\label{mixing-TSS-MWT}
\eeq
where $c_{1,2}$ are dimensionless parameters. This additional potential, Eq.(\ref{mixing-TSS-MWT}), does not contribute to the stationarity conditions, 
determined by the potential Eq.(\ref{TSS-higgs-potential}). 
The vacuum structure of this model is determined by three vacuum expectation values (vevs),  $v_{{\rm TC},1,2}$ , all contributing
to the electroweak symmetry breaking, and satisfying the constraint $v_1^2+v_2^2+v_{\rm TC}^2=v_{\rm EW}^2$, where $v_{\rm EW}=246 \, (\GeV)$. We define $\tan \beta$ and $\tan \phi$ as
\beq
\tan \beta \equiv \frac{v_2}{v_1}
\quad,\quad
\tan^2 \phi \equiv \frac{v^2_{\rm TC}}{v^2_1 + v^2_2}\, ,
\eeq
or in other words,
\beq
\begin{aligned}
& v_{\rm TC} = v_{\rm EW} \sin \phi, \\
& v_1 = v_{\rm EW} \cos \phi \cos \beta, \\
& v_2 = v_{\rm EW} \cos \phi \sin \beta.
\end{aligned}
\label{vev-seesaw}
\eeq

Next, we discuss the higgs boson mass spectrum in the present model. The quadratic terms of the NGB fields arising from the both sectors,$V(\Phi_1,\Phi_2) +V_M(\Phi_1,\Phi_2,\Phi_{\rm TC})$, are given by
\beq
{\cal L}^{\rm higgs}
=
-\frac{1}{2}(\pi^0_1 \,\,\, \pi^0_2 \,\,\, \pi^0_{\rm TC}) {\cal M}^2_{\pi} \bpm \pi^0_1 \\ \pi^0_2 \\ \pi^0_{\rm TC}\epm
-(\pi^+_1 \,\,\, \pi^+_2\,\,\,\pi^+_{\rm TC}) {\cal M}^2_{\pi\pm} \bpm \pi^-_1 \\ \pi^-_2 \\\pi^-_{\rm TC}\epm
-\frac{1}{2}(h^0_1 \,\,\, h^0_2) {\cal M}^2_{h} \bpm h^0_1 \\ h^0_2 \epm .
\label{lag-higgs-mass}
\eeq 
Let us first concentrate for the top-seesaw sector only. Then the CP-odd higgs and charged higgs mass matrices are given by 
\cite{Haber:1993an}
\beq
\left. {\cal M}^2_\pi \right|_{\rm TC = 0}
=
M^2_{12} \bpm \tan \beta & -1 \\ -1 & \tan \beta \epm
\,\label{TSS-CP-odd-higgs-mass}
\eeq
for the CP-odd higgs sector, and
\beq
\left. {\cal M}^2_{\pi \pm} \right|_{\rm TC = 0}
=
\left[ M^2_{12} - \frac{1}{2}  \lambda_4 v^2_{\rm EW} \cos^2 \phi \sin \beta \cos \beta\right]
\bpm \tan \beta & -1 \\ -1 & \tan \beta \epm
\,\label{TSS-charged-higgs-mass}
\eeq
for the charged higgs sector.  It will be convenient to define $M^2_{{\rm TSS},0,\pm}$ as
\beq
M^2_{{\rm TSS},0} = \frac{M^2_{12}}{\cos \beta \sin \beta}
\quad , \quad
M^2_{{\rm TSS},\pm} 
=
M^2_{{\rm TSS},0} - \frac{1}{2} \lambda_4 v^2_{\rm EW} \cos^2\phi\,,
\eeq
which are eigenvalues of Eqs.(\ref{TSS-CP-odd-higgs-mass}) and (\ref{TSS-charged-higgs-mass}). In our study we will treat $M_{{\rm TSS},0}$ as a free parameter. The CP-even higgs boson mass matrices are given by
\beq
{\cal M}^2_h
=
M^2_{{\rm TSS},0}  \bpm \sin^2 \beta & -\sin \beta \cos \beta \\ -\sin \beta \cos \beta & \cos^2 \beta\epm
+
v^2_{\rm EW} \cos^2 \phi
\bpm 2 \lambda_1 \cos^2 \beta & (\lambda_3 + \lambda_4) \sin \beta \cos \beta \\
 (\lambda_3 + \lambda_4) \sin \beta \cos \beta & 2 \lambda_2 \sin^2 \beta \epm
 \,.
 \label{CP-even-higgs-mass}
\eeq
The CP-even higgs boson ($h^0,H^0$) masses, $m_h < m_H$, are eigenvalues of Eq. (\ref{CP-even-higgs-mass}). They are determined solely by the top-seesaw sector, since the TC sector is described by a 
``higgsless" doublet. The mixing angle 
in the CP-even higgs boson sector is defined as
\beq
\tan (2\alpha) = \frac{2 [{\cal M}^2_h]_{12}}{[{\cal M}^2_h]_{11} - [{\cal M}^2_h]_{22}}
\,, \quad
\text{with $-\dfrac{\pi}{2} \leq \alpha \leq 0$}
\,,\label{CP-even-higgs-mixing}
\eeq
and the two CP-even higgs boson mass eigenstates are given by
\beq
\bpm H^0 \\ h^0 \epm
=
\bpm \cos \alpha & \sin \alpha \\ -\sin \alpha & \cos \alpha \epm
\bpm h^0_1 \\ h^0_2 \epm
\,,\label{CP-even-higgs-matrix}
\eeq
which is the same as in the usual 2HDM.  However, we should note the meaning of $\alpha$ in the present model. 
From Eqs. (\ref{reno-yukawa-H4G}) and (\ref{CP-even-higgs-matrix}), we deduce that the two CP-even higgs bosons couple to fermions as 
\beq
[H^0\bar{D}^{(3)}_LD^{(4)}_R\,,\, h^0 \bar{U}^{(3)}_L U^{(4)}_R] \text{-couplings} \propto (\cos \alpha)
\quad , \quad
[h^0\bar{D}^{(3)}_LD^{(4)}_R\,,\, H^0 \bar{U}^{(3)}_L U^{(4)}_R] \text{-couplings} \propto (\sin \alpha)
\,.
\label{CP-even-higgs-constituent}
\eeq
Generally the coupling between the composite higgs and  its constituent fermions is strong.
Therefore, looking at the above couplings, we find that if $|\tan\alpha |<1$, the composite higgs $h^0$ is dominantly a fluctuation of the condensate of up-type quarks. Similarly, if $\tan\alpha |>1$, $h^0$ consists dominantly of a fluctuation around the condensate of down-type quarks. 
Consequently, we can estimate constituent fermion species of the light CP-even higgs boson via the value of $\cos \alpha$. 
%

Then, taking into account the mixing between the top-seesaw sector and TC sector, the mass matrix of the neutral CP-odd 
 higgs boson fields, $\pi^0_i,(i=1,2,{\rm TC})$, including the neutral top-pion of the top-seesaw sector and techni-pion of TC sector, is
\beq
{\cal M}^2_\pi = 
\left(
\begin{array}{cc|c}
&\mbox{\raisebox{-2ex}{\large$\left. {\cal M}^2_\pi \right|_{\rm TC = 0}$}}&0\\
&&0\\ \hline
0&0&0
\end{array}
\right)
+
\bpm
c_1 v^2_1& 0 & - M^2_1 
\\[1ex]
0 &  c_2 v^2_2 & -M^2_2 
\\[1ex]
-M^2_1  & -M^2_2  & M^2_1 \cos \beta \cot \phi + M^2_2 \sin \beta \cot \phi
\epm\,.
\label{mass-NPNGB}
\eeq
Similarly, the mass matrix of charged higgs boson field, $\pi^\pm_i$, which includes the charged top-pion and charged techni-pion, is
\beq
{\cal M}^2_{\pi \pm} = 
\left(
\begin{array}{cc|c}
&\mbox{\raisebox{-2ex}{\large$\left. {\cal M}^2_{\pi \pm} \right|_{\rm TC = 0}$}}&0\\
&&0\\ \hline
0&0&0
\end{array}
\right)
+
\bpm
 c_1 v^2_1& 0& - M^2_1  
\\[1ex]
0&  c_2 v^2_2 & - M^2_2 
\\[1ex]
- M^2_1   & - M^2_2   & M^2_1 \cos \beta \cot \phi + M^2_2 \sin \beta \cot \phi 
\epm\,.
\label{mass-CPNGB}
\eeq
In the above equations, we have defined the mixing mass term between top-seesaw sector and TC sector in $V_M(\Phi_1,\Phi_2,\Phi_{\rm TC})$ as 
\beq
M^2_1 = c_1 v^2_1 \frac{v_1}{v_{\rm TC}}
\quad , \quad
M^2_2 = c_2 v^2_2 \frac{v_2}{v_{\rm TC}}
\,.
\eeq
The CP-odd and charged higgs bosons are represented in terms of the mass basis as
\beq
\bpm G^0 \\[1ex] A^0_2 \\[1ex] A^0_1 \epm
=
O^T_0
\bpm \pi^0_1 \\[1ex] \pi^0_2\\[1ex] \pi^0_{\rm TC} \epm
\quad , \quad 
\bpm G^\pm\\[1ex] H^\pm_2 \\[1ex] H^\pm_1 \epm
=
{\cal O}^T_\pm
\bpm \pi^\pm_1 \\[1ex] \pi^\pm_2\\[1ex] \pi^\pm_{\rm TC} \epm
\,,
\eeq
where the orthogonal matrix $O_p\,\,(p=0,\pm)$ is given as \cite{Hashimoto:2009ty}
\beq
O_p
= \bpm 
\cos \phi \cos \beta & - \sin \beta \cos \zeta_p + \sin \phi \cos \beta \sin \zeta_p &  -\sin \beta \sin \zeta_p - \sin \phi \cos \beta \cos \zeta_p
\\[1ex]
\cos \phi \sin \beta & \cos \beta \cos \zeta_p + \sin \phi \sin \beta \sin \zeta_p & \cos \beta \sin \zeta_p - \sin \phi \sin \beta \cos \zeta_p 
\\[1ex]
\sin \phi & -\cos \phi  \sin \zeta_p & \cos \phi \cos \zeta_p
\epm
\label{ortomatrix-PNGB}
\,.
\eeq
The states $G^{0,\pm}$ become the longitudinal components of the weak gauge bosons and the corresponding  mass eigenvalues are  $M^2_{G^{0,\pm}} = 0$. The non-zero eigenvalues are given as 
\beq
2\hat{M}^2_{S_2}
\!\!&=&\!\!
M^2_{{\rm TSS},p} + 
\frac{(\sin^2 \phi + \cos^2 \beta \cos^2\phi) M^2_1}{\cos \beta \cos \phi \sin \phi} +
\frac{(\sin^2 \phi + \sin^2 \beta \cos^2\phi) M^2_2}{\sin \beta \cos \phi \sin \phi}
\nonumber \\
&& \hspace*{-2ex}
- \left[ 
\begin{aligned}
&
\frac{4}{\cos^2 \phi} \left( M^2_1 \sin \beta - M^2_2 \cos \beta \right)^2
\\
& 
+
\left\{ 
M^2_{{\rm TSS},p} + 
\frac{(\sin^2 \beta \sin^2 \phi - \cos^2 \beta) M^2_1}{\cos \beta \cos \phi \sin \phi} +
\frac{(\cos^2 \beta \sin^2 \phi - \sin^2 \beta) M^2_2}{\sin \beta \cos \phi \sin \phi}
\right\} ^2 
\end{aligned}
\right]^{1/2}\,,
\label{PNGB-heavy-mass}
\\[2ex]
2\hat{M}^2_{S_1}
\!\!&=&\!\!
M^2_{{\rm TSS},p} + 
\frac{(\sin^2 \phi + \cos^2 \beta \cos^2\phi) M^2_1}{\cos \beta \cos \phi \sin \phi} +
\frac{(\sin^2 \phi + \sin^2 \beta \cos^2\phi) M^2_2}{\sin \beta \cos \phi \sin \phi}
\nonumber \\
&& \hspace*{-2ex}
+ \left[ 
\begin{aligned}
&
\frac{4}{\cos^2 \phi} \left( M^2_1 \sin \beta - M^2_2 \cos \beta \right)^2
\\
& 
+
\left\{ 
M^2_{{\rm TSS},p} + 
\frac{(\sin^2 \beta \sin^2 \phi - \cos^2 \beta) M^2_1}{\cos \beta \cos \phi \sin \phi} +
\frac{(\cos^2 \beta \sin^2 \phi - \sin^2 \beta) M^2_2}{\sin \beta \cos \phi \sin \phi}
\right\} ^2 
\end{aligned}
\right]^{1/2}
\label{PNGB-light-mass}
\,,
\eeq
where $S=A^0,H^\pm$ and $p=0,\pm$, respectively. 
The mixing angle $\tan \zeta_p$ is defined as 
\beq
\tan \zeta_p
=
\frac{\hat{M}^2_{S_2} \cos \phi  \sin \phi- \left( M^2_1 \cos \beta+ M^2_2 \sin \beta \right) }{\sin \phi \left( M^2_1 \sin\beta -  M^2_2 \cos \beta \right) }\,.
\eeq


To obtain more insight into this spectrum, we briefly consider the case with $c_1 = c_2 = 0$. This basically corresponds to the case 
studied in \cite{Fukano:2012qx}. In this case, $\bpm G & S_2 & S_1 \epm^T$ becomes 
\beq
\bpm G \\[1ex] S_2 \\[1ex] S_1 \epm
=
\bpm
\cos \phi \cos \beta & \cos \phi \sin \beta&  \sin \phi
\\[1ex]
\sin \phi \cos \beta  & \sin \phi \sin \beta  & -\cos \phi 
\\[1ex]
-\sin \beta  & \cos \beta  & 0
\epm
\bpm \pi_1 \\[1ex] \pi_2\\[1ex] \pi_{\rm TC} \epm
\,.\label{PNGB-limit}
\eeq
From Eqs.(\ref{PNGB-heavy-mass}), (\ref{PNGB-light-mass}) and (\ref{PNGB-limit}) with $c_1 = c_2 = 0$ one can easily see that $S_1$, not 
$S_2$, corresponds to the CP-odd higgs bosons in the 2HDM with $M^2_S = M^2_{\text{TSS},p}$. On the other hand the $S_2$ becomes 
massless. To resolve this, we note that TC sector contributes to $\pi_{\rm TC}$ through ETC interactions. 
Hence, on the effective theory level, we should add a mass term for $\pi_{\rm TC}$,
\beq
{\cal L}^{\rm mass}_{\rm ETC} 
=
- m^2_{\rm ETC} \left[ \frac{1}{2}\pi^0_{\rm TC} \pi^0_{\rm TC}+ \pi^+_{\rm TC} \pi^-_{\rm TC} \right]
\label{TC-contribution-PNGB}
\,,
\eeq
with the value of $m^2_{\rm ETC}$ larger than the difference of the mass eigenvalues give by Eqs. (\ref{PNGB-heavy-mass}) and (\ref{PNGB-light-mass}). This will give a large contribution to mass of $S_2$
 but a negligible contribution to mass of $S_1$. 
In other words, we arrange the spectrum so that
mass squared of the state $S_1$ is given by Eq. (\ref{PNGB-light-mass}), while the mass squared of $S_2$ is given by the sum of Eq. (\ref{PNGB-heavy-mass}) and $m^2_{\rm ETC}$. The Goldstone boson $G$ which is absorbed by the electroweak gauge boson of course remains massless.
Evidently, we do not know $m_{\rm ETC}$ quantitatively unless we consider a concrete ETC model. In this paper we will set $m_{\rm ETC} =  \Lambda_{\rm TC} = 4 \pi v_{\rm TC}$ corresponding to the cutoff scale for the non-linear sigma model which we use to describe the TC sector. 


%
\section{Renormalization group equations and the compositeness conditions}
\label{dynamical-results}
%

\begin{figure}[htbp]
\begin{center}
\begin{tabular}{cc}
{
\begin{minipage}[t]{0.5\textwidth}
\begin{flushleft} (a) \end{flushleft} \vspace*{-5ex}
\includegraphics[scale=0.73]{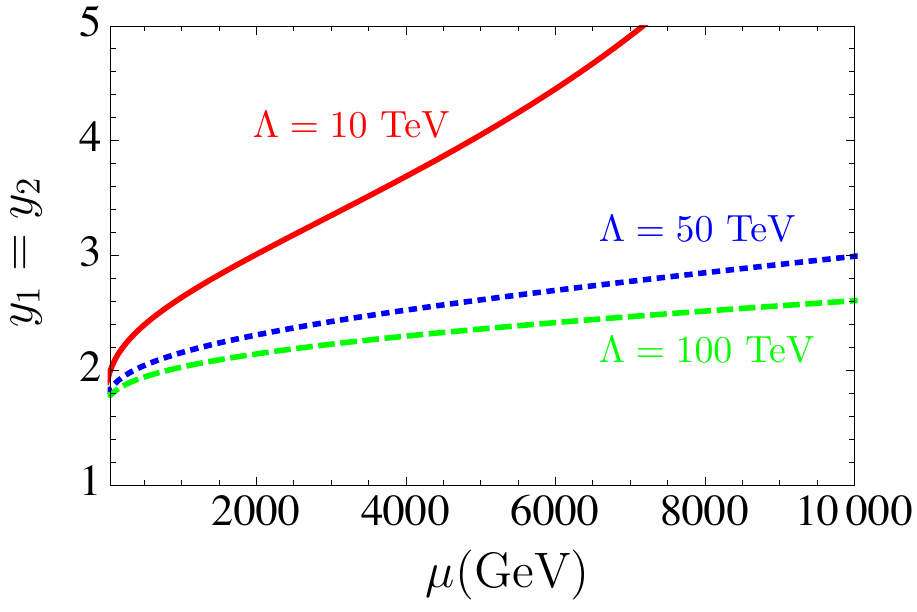} 
\vspace*{2ex}
\end{minipage}
}
{
\begin{minipage}[t]{0.5\textwidth}
\begin{flushleft} (b) \end{flushleft} \vspace*{-5ex}
\includegraphics[scale=0.8]{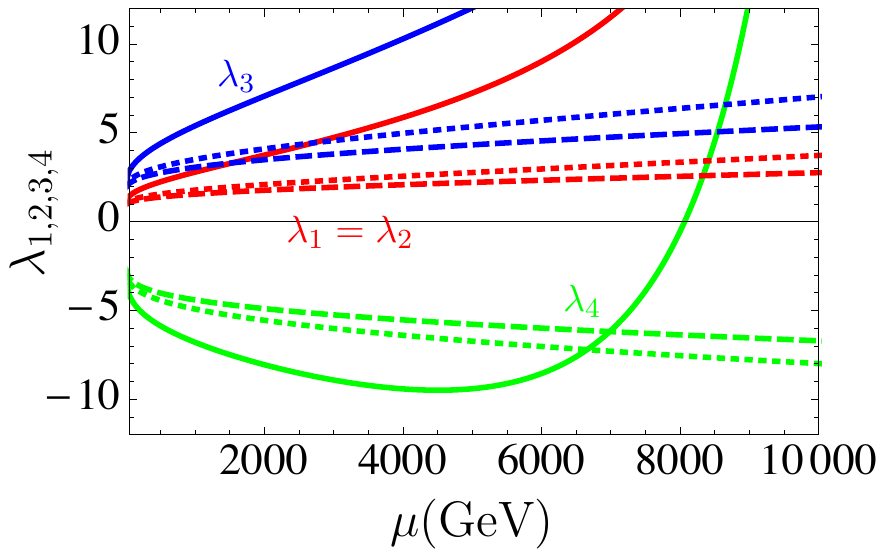} 
\end{minipage}
}
\end{tabular}
\caption[]{
The scale dependence of (a) yukawa couplings and (b) quartic couplings. In each panel, the solid, dotted and dashed lines correspond to $\Lambda = 10,50$ and $100 \TeV$, respectively. In panel (b), the red curves correspond to $\lambda_1=\lambda_2$, and blue and green curves correspond to 
$\lambda_3$ and $\lambda_4$, respectively.
\label{RGE-CC}}
\end{center}
\end{figure}

In the previous paper \cite{Fukano:2012qx}, we analyzed the dynamics by using the gap equations. 
In order to carry out more precise analysis 
in this paper, we study the model using the renormalization group 
equations  (RGEs) together with compositness conditions \cite{Bardeen:1989ds}. 
We ignore the RGEs of SM electroweak interaction since 
their contributions are negligible at the relevant energy scales, and
consider only the RGEs for QCD gauge coupling, 
Yukawa couplings and the higgs quartic couplings. 
The RGE for $SU(3)_c$ gauge coupling is given by
\beq
(16 \pi^2) \mu \frac{d g_3}{d \mu}
=
-\left[ 11 - \frac{4}{3} N_g\right]g^3_3
\,,\label{RGE-QCD}
\eeq 
with the initial condition $\alpha_{\rm QCD}(M^2_Z) \equiv g^2_3(M^2_Z)/(4 \pi^2) = 0.1184$. Here $N_g$ is number of fermion generation which is $N_g = 4$ in the present model.
The RGEs for yukawa couplings $y_{1,2}$ in Eq.(\ref{reno-yukawa-H4G}) are given by
\beq
(16 \pi^2) \mu \frac{d y_1}{d \mu}
\!\!\!&=&\!\!\!
\left[ - 8 g^2_3 + \frac{9}{2} y^2_1 + \frac{1}{2} y^2_2 \right] y_1 
\label{RGE-y1}
\,,\\[1ex]
(16 \pi^2) \mu \frac{d y_2}{d \mu}
\!\!\!&=&\!\!\!
\left[ - 8 g^2_3 + \frac{9}{2} y^2_2 + \frac{1}{2} y^2_1 \right] y_2 
\,,
\label{RGE-y2}
\eeq
and the RGEs for higgs quartic couplings $\lambda_{1,2,3,4}$ for the top-seesaw sector in Eq.(\ref{TSS-higgs-potential}) are given by \cite{Branco:2011iw,Haber:1993an}
\beq
(16 \pi^2) \mu \frac{d \lambda_1}{d \mu}
\!\!\!&=&\!\!\!
24 \lambda^2_1 + 2 \lambda^2_3 + 2 \lambda_3 \lambda_4 + \lambda^2_4 
+ 12 \lambda_1 y^2_1 - 6y^4_1
\,,\label{RGE-lambda1}\\[1ex]
(16 \pi^2) \mu \frac{d \lambda_2}{d \mu}
\!\!\!&=&\!\!\!
24 \lambda^2_2 + 2 \lambda^2_3 + 2 \lambda_3 \lambda_4 + \lambda^2_4 
+ 12 \lambda_2 y^2_2 - 6y^4_2
\,,\label{RGE-lambda2}\\[1ex]
(16 \pi^2) \mu \frac{d \lambda_3}{d \mu}
\!\!\!&=&\!\!\!
2( \lambda_1+ \lambda_2)(6 \lambda_3 + 2 \lambda_4) 
+ 4 \lambda^2_3 + 2 \lambda^2_4 + 6 \lambda_3 (y^2_1 + y^2_2) - 12 y^2_1 y^2_2
\,,\label{RGE-lambda3}\\[1ex]
(16 \pi^2) \mu \frac{d \lambda_4}{d \mu}
\!\!\!&=&\!\!\!
4( \lambda_1+ \lambda_2)  \lambda_4 
+ 4 (2 \lambda_3 +  \lambda_4) \lambda_4 + 6 \lambda_4 (y^2_1 + y^2_2) + 12 y^2_1 y^2_2
\,.\label{RGE-lambda4}
\eeq
The compositeness conditions in this model are given by \cite{Bardeen:1989ds,Hashimoto:2009ty}
\beq
&&
y^2_{1,2}(\mu) \to y^2_{1,2}(\Lambda) = \infty
\label{cc-yukawa}
\\[1ex]
&&
\frac{\lambda_1(\mu)}{y^4_1(\mu)} \to   \frac{\lambda_1(\Lambda)}{y^4_1(\Lambda)} =0
\quad , \quad
\frac{\lambda_2(\mu)}{y^4_2(\mu)} \to   \frac{\lambda_2(\Lambda)}{y^4_2(\Lambda)} =0,
\label{cc-lambda12}
\\[1ex]
&&
\frac{\lambda_3(\mu)}{y^2_1(\mu)y^2_2(\mu)} \to   \frac{\lambda_3(\Lambda)}{y^2_1(\Lambda)y^2_2(\Lambda)} =0
\quad , \quad
\frac{\lambda_4(\mu)}{y^2_1(\mu)y^2_2(\mu)} \to   \frac{\lambda_4(\Lambda)}{y^2_1(\Lambda)y^2_2(\Lambda)} =0\,,
\label{cc-lambda34}
\eeq
where $\Lambda$ is called a compositeness scale and this scale is identified with the mass scale of 
the massive topcolor gluons $M_{G'}$ in the present model. The dynamics is then determined as follows: As a first step, we solve system of RGEs, Eqs.(\ref{RGE-QCD})
-(\ref{RGE-lambda4}), under the compositeness conditions, Eqs.(\ref{cc-yukawa})%
-(\ref{cc-lambda34}) for given $\Lambda$. As a second step, we find the physical solutions for $y_{1,2}$
and $\lambda_{1,2,3,4}$ from the on-shell conditions
\beq
&&
\Sigma_{D} = \frac{ v_1}{\sqrt{2}} \,y_1(\mu = \Sigma_D)
\quad ,\quad
\Sigma_{U} = \frac{ v_2}{\sqrt{2}} \,y_2(\mu = \Sigma_U) 
\,,\label{on-shell-condition-tb}
\\[1ex]
&&
m_H = 
m_H (M_{\rm TSS,0},\lambda_1(m_H),\lambda_2(m_H),\lambda_3(m_H),\lambda_4(m_H))
\,,\label{on-shell-condition-higgs}
\eeq
where $\Sigma_{U,D}$ is the dynamical fermion mass and $m_H$ is the heavy CP-even higgs boson mass. 
In Fig.\ref{RGE-CC}, we show the results of solving the RGEs with the compositeness conditions for $\Lambda =10,50,100 \TeV$. This corresponds to the first step described above. 
%
%
%
%

%
%

\begin{figure}[htbp]
\begin{center}
\begin{tabular}{cc}
{
\begin{minipage}[t]{0.5\textwidth}
\includegraphics[scale=0.8]{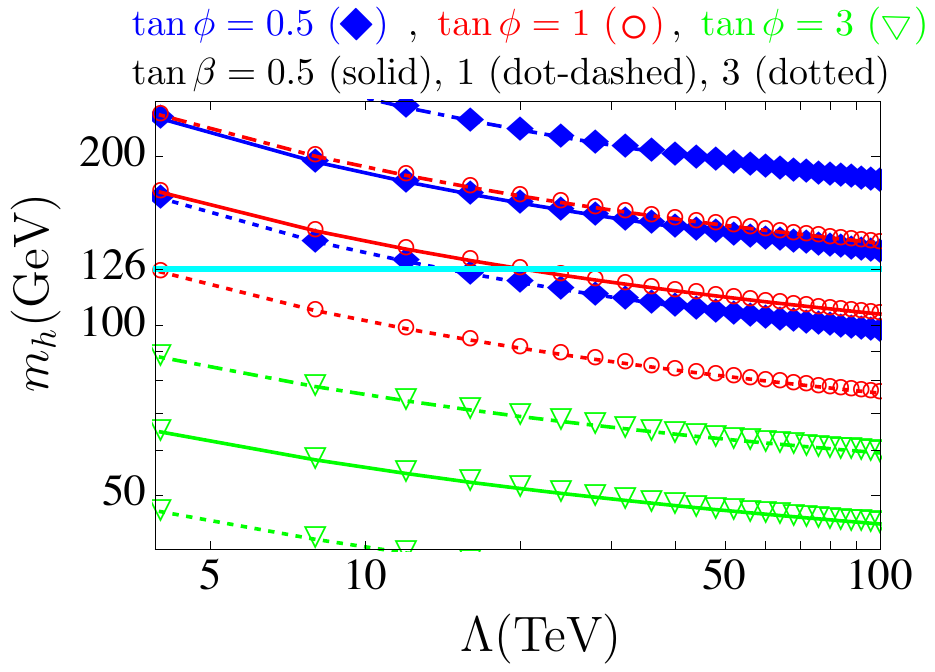} 
\vspace*{2ex}
\end{minipage}
}
\end{tabular}
\caption[]{
The dynamical higgs mass in the top-seesaw sector for $4 \TeV \leq \Lambda \leq 100 \TeV$, i.e. $m_h$ for Eq.(\ref{CP-even-higgs-mass}) with $M_{\rm{TSS},0} = 0$. The horizontal cyan solid line shows $m_h = 126 \GeV$.
\label{possible-126GeVhiggs}}
\end{center}
\end{figure}

We want to see if a light CP-even higgs with mass  around $126 \GeV$ 
can be accommodated within the model 
for arbitrary $\Lambda$ with $ \Lambda\geq 4 \TeV $ which is satisfied with the 
lower bound $M_{G'} > 3.32 \TeV$ at $95 \%\cl$ from LHC \cite{Aad:2011fq}.
First, we consider $M_{{\rm TSS},0} = 0$ in Eqs.(\ref{CP-even-higgs-mass}) and 
(\ref{on-shell-condition-higgs}), and $c_1 = c_2 = 0$ in Eq.(\ref{mixing-TSS-MWT}).  
We solve the systems of RGEs,  Eqs.(\ref{RGE-QCD})
-(\ref{RGE-lambda4}) under the compositeness conditions, Eqs.(\ref{cc-yukawa})
-(\ref{cc-lambda34}) with the on-shell condition Eq.(\ref{on-shell-condition-higgs}). 
%
In Fig. \ref{possible-126GeVhiggs}, we show the resultant $m_h$ together with $m_h = 126 \GeV$ line (horizontal cyan solid line) for $\tan \phi = 0.5,1,3$ and $\tan \beta = 0.5,1,3$. For nonzero and positive values of $M_{{\rm TSS},0} $, the scalar mass $m_h$ should be below the $m_h = 126 \GeV$ line, and we immediately find that the values $(\tan \phi , \tan \beta) = (0.5,0.5),(0.5,1),(1,1)$ are disfavored for $m_h = 126 \GeV$.

\begin{figure}[htbp]
\begin{center}
\begin{tabular}{cc}
{
\begin{minipage}[t]{0.5\textwidth}
\begin{flushleft} (a) \end{flushleft} \vspace*{-5ex}
\includegraphics[scale=0.75]{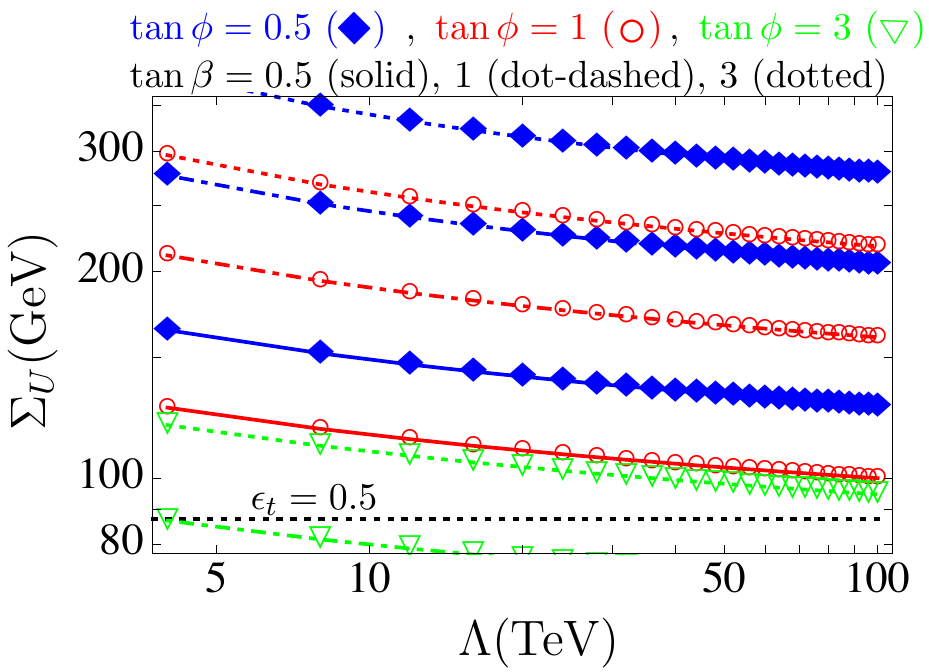} 
\vspace*{2ex}
\end{minipage}
}
{
\begin{minipage}[t]{0.5\textwidth}
\begin{flushleft} (b) \end{flushleft} \vspace*{-5ex}
\includegraphics[scale=0.75]{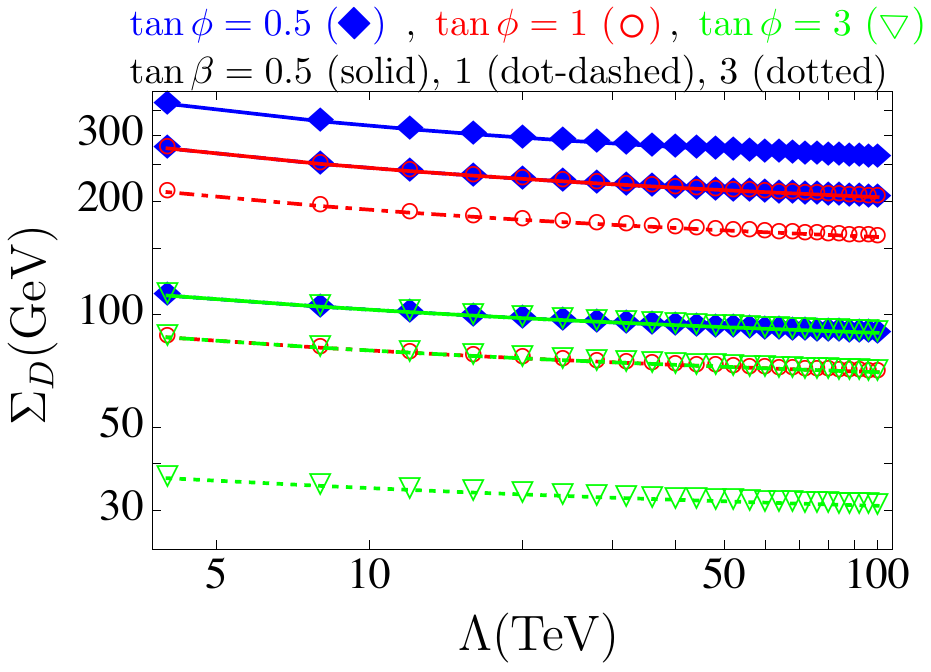} 
\end{minipage}
}
\end{tabular}
\caption[]{
The dynamical fermion mass for (a) top sector and (b) bottom sector with $\tan \phi  = 0.5,1,3$ and $\tan \beta = 0.5,1,3$. The horizontal dotted line in (a) corresponds to  $m_t(\text{TSS})=0.5 m_t$
. 
\label{fermion-dynamical-mass}}
\end{center}
\end{figure}

To constrain the allowed values of $(\tan\phi\,,\,\tan \beta)$ further, we consider the fermion masses. Since the top and bottom quark masses are sourced from ETC interactions as well as from the top-seesaw sector, we take them to be represented as 
\beq
m_t 
&=& m_t({\rm ETC}) + m_t({\rm TSS}) = \epsilon_t m_t + (1-\epsilon_t) m_t
\,,\label{physical-top-mass}\\[1ex]
m_b 
&=&  m_b({\rm ETC}) + m_b({\rm TSS})= \epsilon_b m_b + (1-\epsilon_b) m_b
\,.\label{physical-bottom-mass}
\eeq
Here $m_{t,b}({\rm ETC}) \equiv  \epsilon_{t,b} m_{t,b} $ and 
$m_{t,b}({\rm TSS}) \equiv  (1 - \epsilon_{t,b}) m_{t,b} $ correspond to 
the contributions to the top/bottom quark mass arising from the four fermion interactions
due to the ETC sector and the top-seesaw sector, respectively. 
In the spirit of the original top-seesaw 
model, we require $0 \leq \epsilon_t \leq 0.5$ 
corresponding to $m_t({\rm ETC}) \leq m_t({\rm TSS})$.
To begin with, we fix
\beq
\epsilon_t = \epsilon_b = 0.5\,,
\label{sample-epsilontb}
\eeq
as representative values. In Fig. \ref{fermion-dynamical-mass}, we show the dynamical fermion mass 
$\Sigma_{U/D}$ for $4 \TeV \leq \Lambda \leq 100 \TeV$. In this figure, we take $\tan \phi  = 0.5,1,3$ 
and $\tan \beta = 0.5,1,3$ as in Fig.\ref{possible-126GeVhiggs}.  In order to realize the 
top-seesaw dynamics, we must have $\Sigma_U > m_t({\rm{TSS}}) = (1-\epsilon_t) m_t$ 
with $\epsilon_t = 0.5$. This limiting value of $m_t(\rm{TSS})$ in the case $\epsilon_t=0.5$ is
shown  as the horizontal dotted line in the left panel of Fig.\ref{fermion-dynamical-mass}. 
On the other hand, for the bottom sector, $\Sigma_D$ is always larger than 
$m_b \simeq 4 \GeV$, so no additional constraints arise here. Thus, combining 
Fig.\ref{possible-126GeVhiggs} and Fig.\ref{fermion-dynamical-mass}, we take the 
benchmark values of $(\epsilon_t, \tan \phi,\tan\beta)$ as $\epsilon_t = 0.5$ and 
\begin{align}
 \tan \phi = 1 \quad,&\quad \tan \beta =0.5 \,, 
 \label{tptb-sample0}
 \\[1ex]
\tan \phi = 1\quad,&\quad \tan \beta =3 \,, 
 \label{tptb-sample1}
\\[1ex]
 \tan \phi = 0.5\quad,&\quad \tan \beta =3 \,, 
 \label{tptb-sample2}
\\[1ex]
 \tan \phi = 3\quad,&\quad \tan \beta =3 \,.
 \label{tptb-sample3}
 \end{align}

%

\begin{figure}[htbp]
\begin{center}
\begin{tabular}{cc}
{
\begin{minipage}[t]{0.5\textwidth}
\begin{flushleft} (a) \end{flushleft} \vspace*{-5ex}
\includegraphics[scale=0.72]{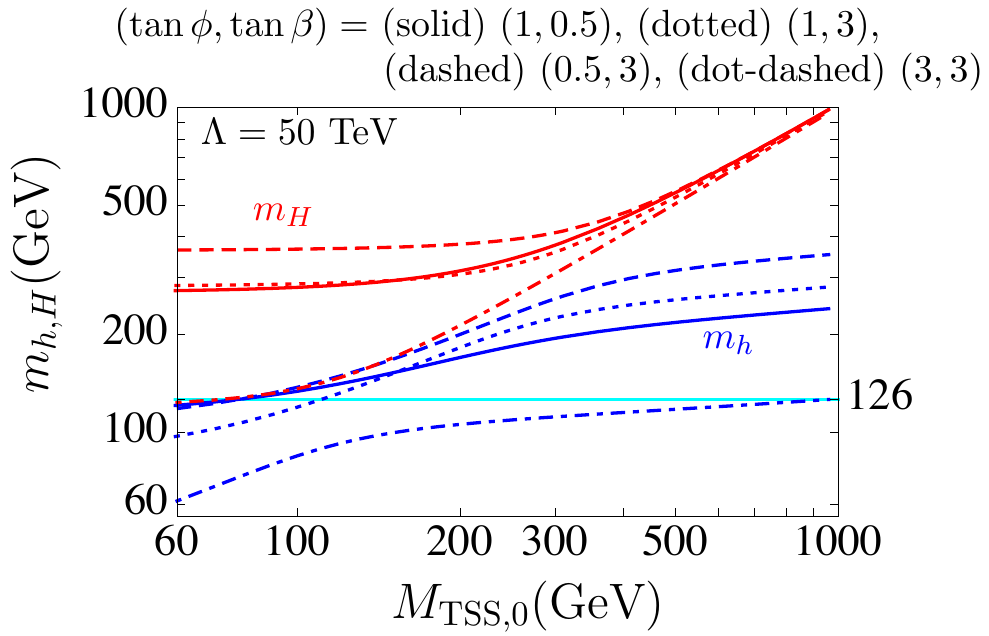} 
\vspace*{2ex}
\end{minipage}
}
{
\begin{minipage}[t]{0.5\textwidth}
\begin{flushleft} (b) \end{flushleft} \vspace*{-5ex}
\includegraphics[scale=0.7]{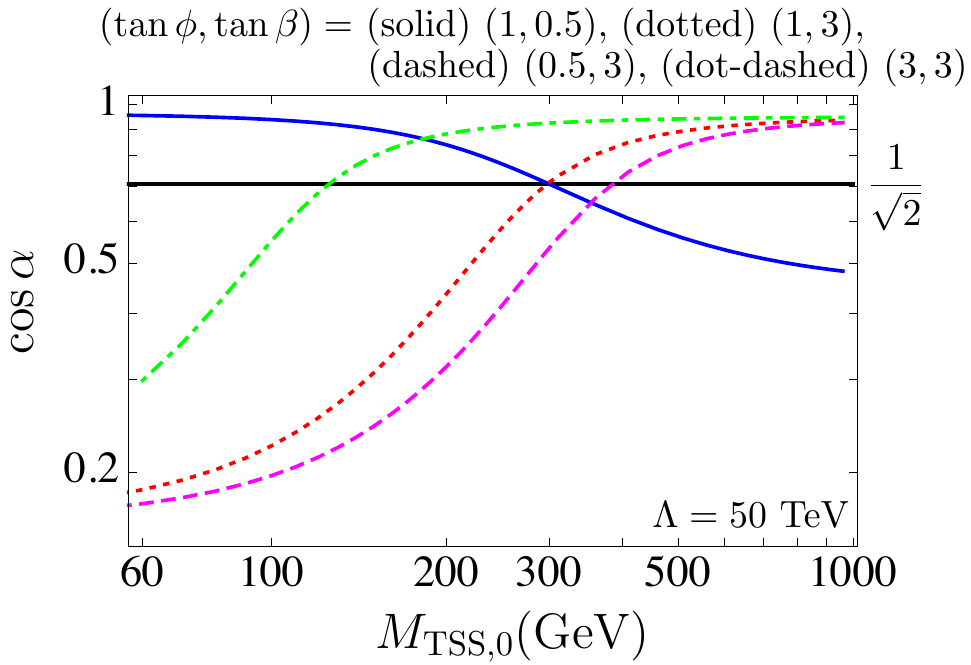} 
\vspace*{2ex}
\end{minipage}
}
\end{tabular}
\caption[]{
(a) The CP-even higgs boson masses and (b) the mixing angles ($\alpha$ in Eq.(\ref{CP-even-higgs-mixing}))  between two CP-even higgs  bosons at $\Lambda = 50 \TeV$ with varying $M_{{\rm TSS},0}$.  In both panels, each curve corresponds to $(\tan \phi\,,\,\tan \beta) =$ (solid) $(1,0.5)$, (dotted) $(1,3)$, (dashed) $(0.5,3)$ and (dot-dashed) $(3,3)$. The horizontal cyan solid line shows (a) $m_h = 126 \GeV$ and (b) $\alpha = \pi/4$, i.e. $|\cos \alpha| = |\sin \alpha|$.
\label{CP-even-higgs}}
\end{center}
\end{figure}

%
We focus mainly on $\Lambda = 50 \TeV$, which also diminishes the contributions from the massive topcolor gauge bosons to the electroweak precision parameters \cite{Fukano:2012qx,He:2001fz}. %
To fix the parameter $M_{{\rm TSS},0}$, we consider the benchmark parameter values listed above, $\epsilon_t=0.5$ and $\Lambda = 50 \TeV$. For each of these parameter sets, 
in Fig. \ref{CP-even-higgs}, we show (a) the CP-even higgs boson masses and (b) their mixing 
angles. 
From Fig. \ref{CP-even-higgs}(a), 
we find that for the lighter state, $h$, the value $m_h = 126 \GeV$ can be realized if $M_{{\rm TSS},0} \simeq 100 \GeV$ for cases in Eqs.(\ref{tptb-sample0},\ref{tptb-sample1},\ref{tptb-sample2}) or $M_{{\rm TSS},0} \simeq 1 \TeV$ for the 
case in Eq.(\ref{tptb-sample3}). 
For the case of Eq.(\ref{tptb-sample3}), corresponding to the dot-dashed curves 
in Fig.\ref{CP-even-higgs}(a), it is also possible to have the heavier state, $H$, to satisfy 
$m_H =126 \GeV$ for $M_{{\rm TSS},0} \simeq 70 \GeV$. We will return to this special case 
shortly, but consider first the case of the lighter state $h$ satisfying $m_h = 126 \GeV$. 
In Fig.\ref{CP-even-higgs}(b), the horizontal solid line is $\cos \alpha = 1/\sqrt{2}$, and above 
(below) this line  $|\tan \alpha| < 1$ $(|\tan \alpha| > 1)$. Based on Fig. \ref{CP-even-higgs} and the 
discussion below Eq.  (\ref{CP-even-higgs-constituent}), we expect that the  state $h$ with mass of 
126 GeV originates mainly from the condensate $\vev{\overline{U}^{(3)}_L U^{(4)}_R} \neq 0$ 
in the case of Eqs. (\ref{tptb-sample0}) and (\ref{tptb-sample3}). On the other hand,
in the case of Eqs. (\ref{tptb-sample1}) and (\ref{tptb-sample2})  the state $h$, and in the case of 
Eq. (\ref{tptb-sample3}) the state $H$, come mainly from $\vev{\overline{D}^{(3)}_L D^{(4)}_R} \neq 0$. Therefore, in 
order to realize $m_{h,H} = 126 \GeV$ at $\Lambda = 50 \TeV$ we take 
\beq
M_{{\rm TSS},0} = 77 \GeV && \text{for} \quad \tan \phi = 1 \quad , \quad \tan \beta =0.5 
\,, \label{MTSS0-sample0}
\\[1ex]
M_{{\rm TSS},0} = 111 \GeV  \!\!\!&&\text{for} \quad \tan \phi = 1 \quad , \quad \tan \beta =3 
\,, \label{MTSS0-sample1}
\\[1ex]
M_{{\rm TSS},0} = 78 \GeV && \text{for} \quad \tan \phi = 0.5 \quad\!\!\! , \quad \!\!\tan \beta =3 \,. \label{MTSS0-sample2}
\\[1ex]
M_{{\rm TSS},0} = 960 \GeV \!\!\!&& \text{for} \quad \tan \phi = 3 \quad , \quad \tan \beta =3 \,, \label{MTSS0-sample3}
\eeq
for $m_h = 126 \GeV$ and 
\beq
M_{{\rm TSS},0} = 73 \GeV \!\!\!&& \text{for} \quad \tan \phi = 3 \quad , \quad \tan \beta =3 \,, \label{MTSS0-sample4}
\eeq
for $m_H = 126 \GeV$.

\begin{figure}[htbp]
\begin{center}
\begin{tabular}{cc}
{
\begin{minipage}[t]{0.5\textwidth}
\begin{flushleft} (a-1) \end{flushleft} \vspace*{-5ex}
\includegraphics[scale=0.65]{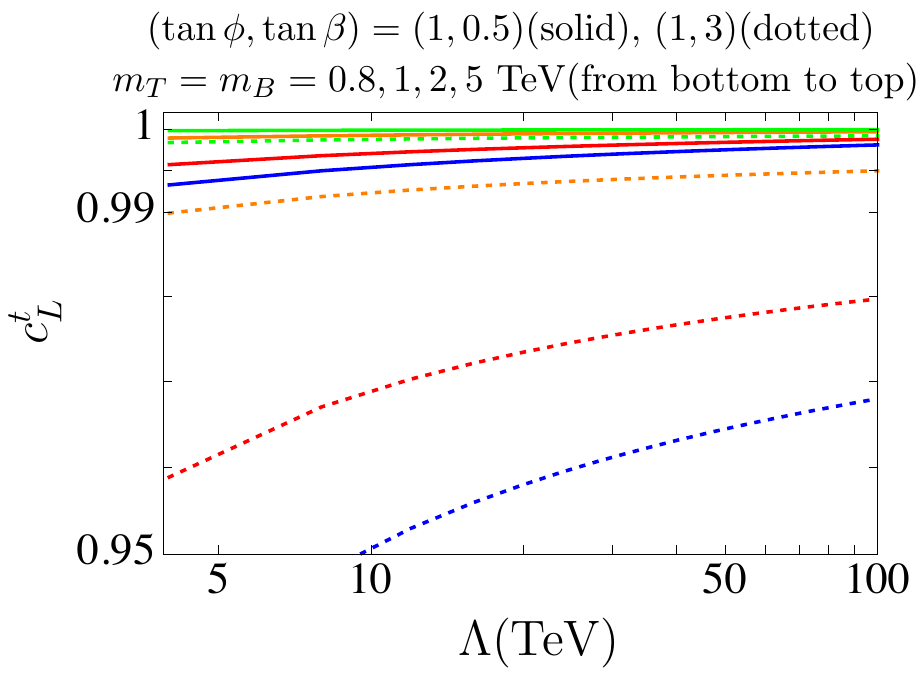} 
\vspace*{1ex}
\end{minipage}
}
{
\begin{minipage}[t]{0.5\textwidth}
\begin{flushleft} (a-2) \end{flushleft} \vspace*{-5ex}
\includegraphics[scale=0.65]{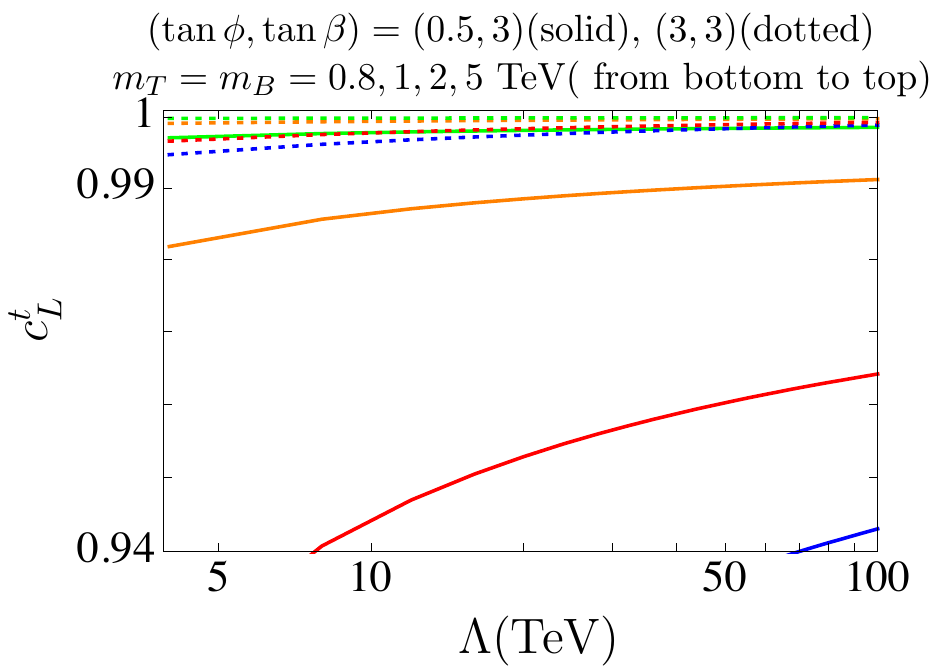} 
\end{minipage}
}
\\
{
\begin{minipage}[t]{0.5\textwidth}
\begin{flushleft} (b-1) \end{flushleft} \vspace*{-5ex}
\includegraphics[scale=0.65]{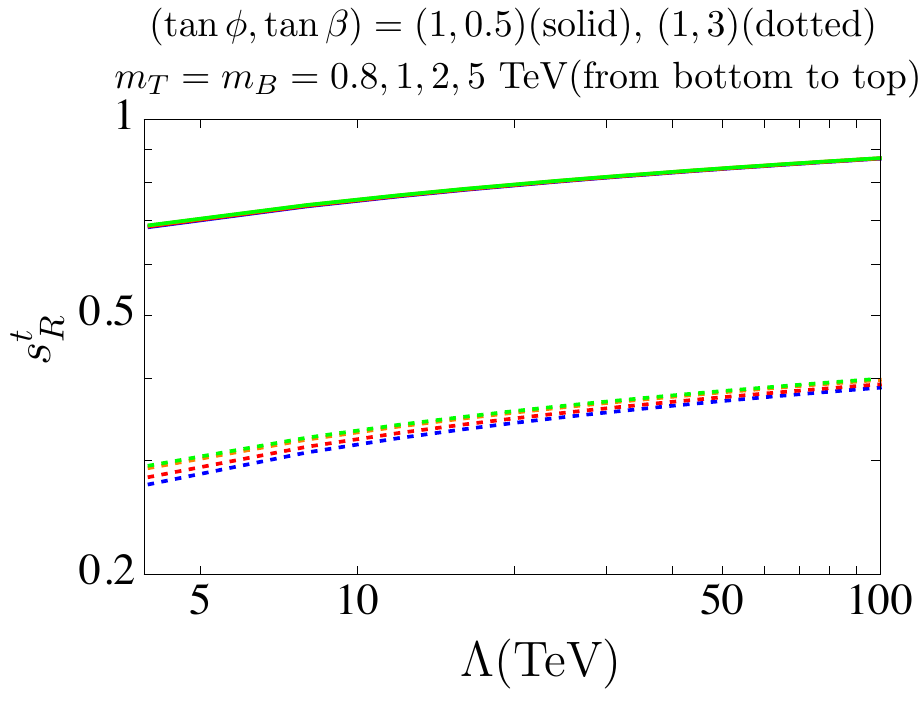} 
\vspace*{1ex}
\end{minipage}
}
{
\begin{minipage}[t]{0.5\textwidth}
\begin{flushleft} (b-2) \end{flushleft} \vspace*{-5ex}
\includegraphics[scale=0.65]{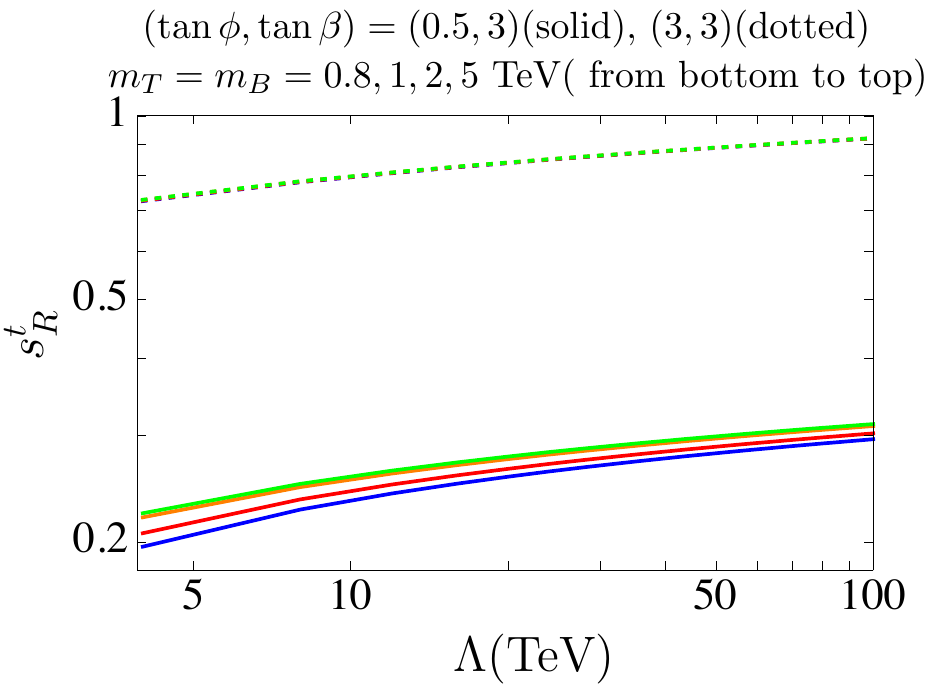} 
\end{minipage}
}
\\
{
\begin{minipage}[t]{0.5\textwidth}
\begin{flushleft} (c-1) \end{flushleft} \vspace*{-5ex}
\includegraphics[scale=0.65]{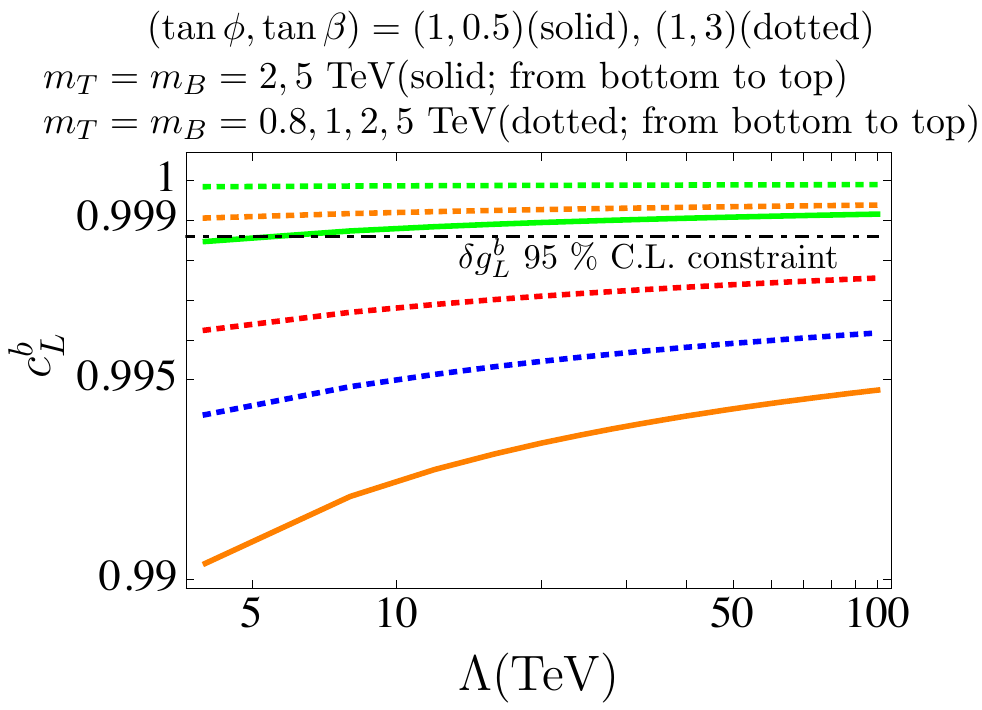} 
\vspace*{1ex}
\end{minipage}
}
{
\begin{minipage}[t]{0.5\textwidth}
\begin{flushleft} (c-2) \end{flushleft} \vspace*{-5ex}
\includegraphics[scale=0.65]{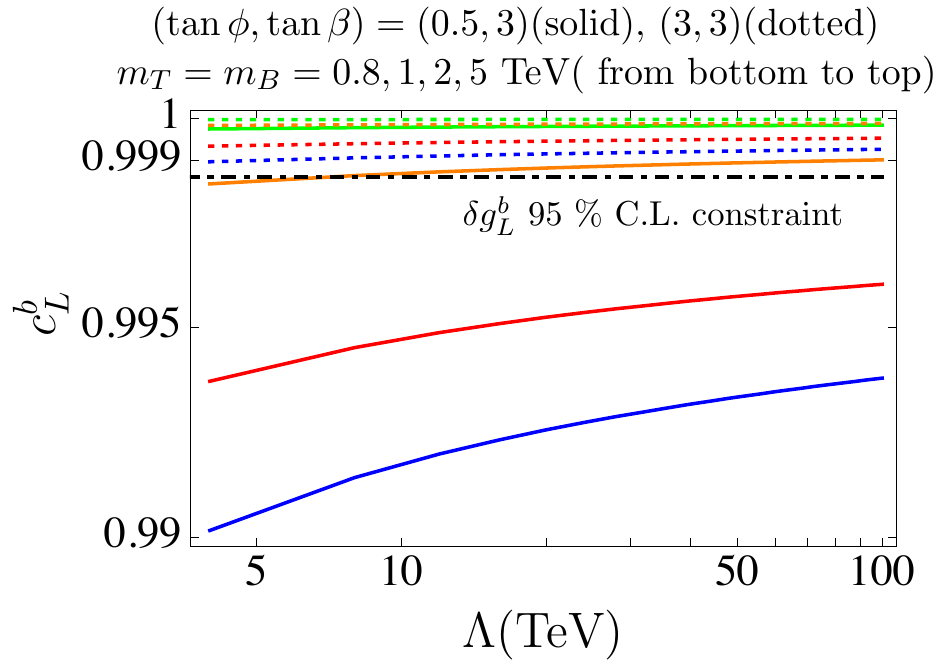} 
\end{minipage}
}
\\
{
\begin{minipage}[t]{0.5\textwidth}
\begin{flushleft} (d-1) \end{flushleft} \vspace*{-5ex}
\includegraphics[scale=0.65]{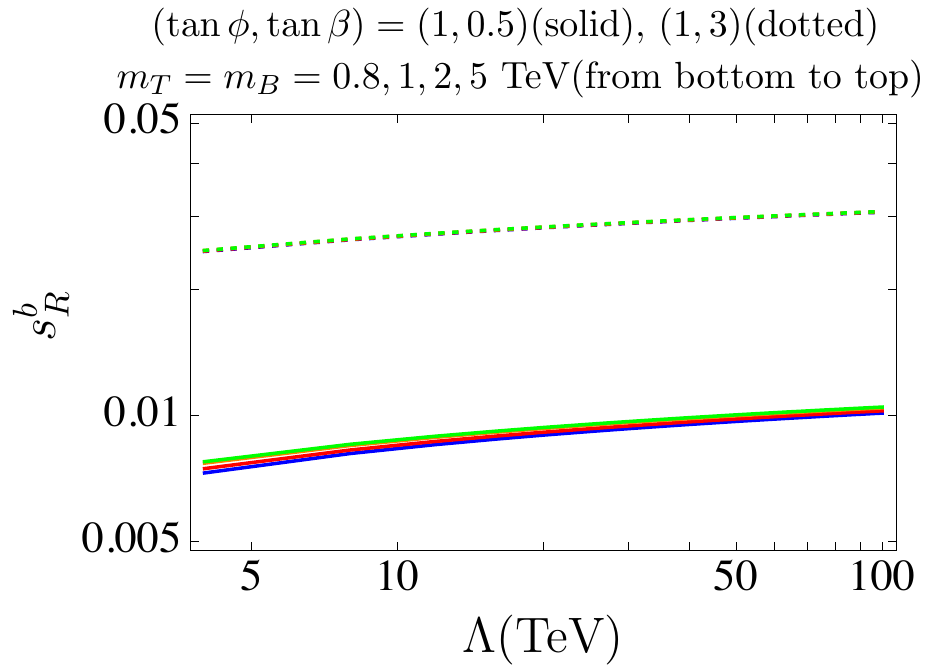} 
\vspace*{1ex}
\end{minipage}
}
{
\begin{minipage}[t]{0.5\textwidth}
\begin{flushleft} (d-2) \end{flushleft} \vspace*{-5ex}
\includegraphics[scale=0.65]{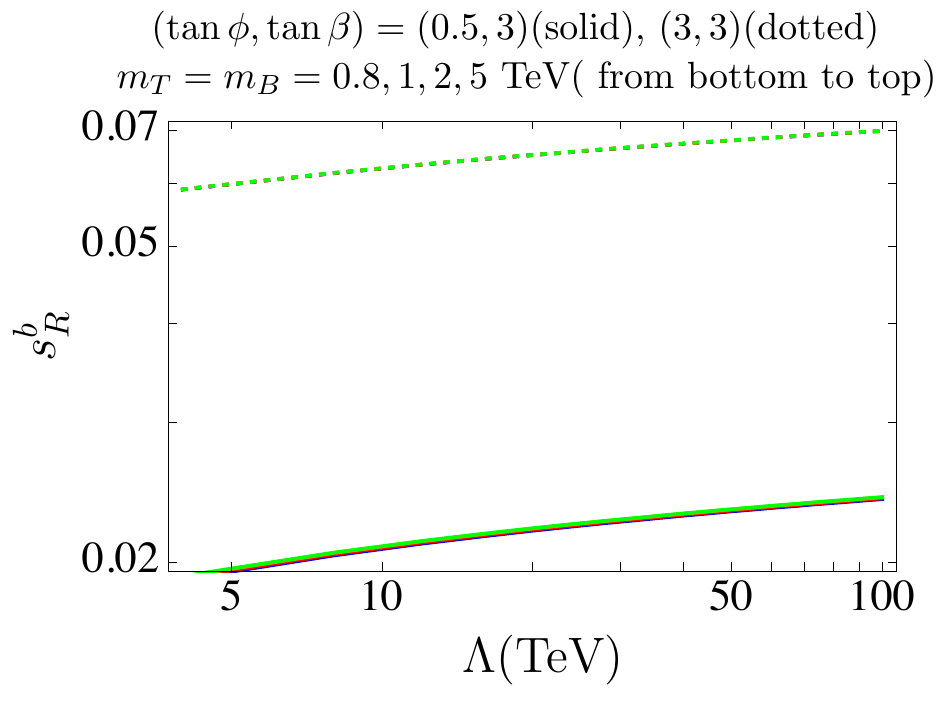} 
\end{minipage}
}
\end{tabular}
\caption[]{
The mixing angles of fermions. The panels (a,b,c,d-1) correspond to $(\tan \phi\,,\,\tan \beta) =$ (solid curves) $(1,0.5)$ and (dotted curves) $(1,3)$, and (a,b,c,d-2) correspond to $(\tan \phi\,,\,\tan \beta) =$ (solid curves) $(0.5,3)$ and (dotted curves) $(3,3)$. In all panels, $m_{T,B} = 800 \GeV, 1,2,5 \TeV$ from bottom curves to top curves (blue,red,orange,green)  in the same $(\tan \phi\,,\,\tan \beta)$-groups. The horizontal magenta dot-dashed line in (c-1,2) shows the $95 \% \cl $ allowed line by the constraint for $\delta g^b_L$ (see section.\ref{section-Zbb}).
\label{fermion-mixing-angle}}
\end{center}
\end{figure}

Based on the benchmark parameters, Eqs. (\ref{tptb-sample0})-
(\ref{tptb-sample3}), we next discuss the quark mixing angles. First, the fermion mass part after the dynamical electroweak symmetry breaking is 
\beq
-
\bpm \bar{U}^{(3)}_L & \bar{U}^{(4)}_L\epm
\bpm 0 & \Sigma_U \\[1ex] M^{(43)}_{U} & M^{(44)}_U\epm
\bpm U^{(3)}_R \\[1ex] U^{(4)}_R\epm
-
\bpm \bar{D}^{(3)}_L & \bar{D}^{(4)}_L\epm
\bpm 0 & \Sigma_D \\[1ex] M^{(43)}_{D} & M^{(44)}_D\epm
\bpm D^{(3)}_R \\[1ex] D^{(4)}_R\epm
+ {\rm h.c.}
\,,
\label{seesaw-mass-part}
\eeq
where $M^{(43,44)}_{U,D}$ are the mass parameters which do not contribute to the dynamical electroweak symmetry breaking, and they are arbitrary parameters in the present model framework. 
%
%
Now, we assume that the quark mixing matrices $U, D$ reflect the seesaw mechanism for the third generation and their vector-like partners, and hence the quark mixing matrices are written as
\beq
U^L_{\alpha\beta}\simeq
\bpm 
1 & 0 & 0 & 0 \\
0 & 1 & 0 & 0 \\
0 & 0 & c^t_L & s^t_L \\
0 & 0 & -s^t_L & c^t_L
\epm
\quad &,& \quad 
U^R_{\alpha\beta}
\simeq
\bpm 
1 & 0 & 0 & 0 \\
0 & 1 & 0 & 0 \\
0 & 0 & -c^t_R & s^t_R \\
0 & 0 & s^t_R & c^t_R
\epm\,,
\,\label{seesaw-mixing-u}
\\[2ex]
D^L_{\alpha\beta} =U^L_{\alpha\beta}\left.\right|_{t \to b}
\quad \hspace*{8ex} &,& \quad
D^R_{\alpha\beta} = U^R_{\alpha\beta}\left.\right|_{t \to b}\,.
\label{seesaw-mixing-d}
\eeq
where $c^t_L \equiv \cos \theta^t_L\,,\,s^t_R \equiv \sin \theta^t_R$, etc. These fermion mixing 
matrices, $U$ and $D$ in Eqs.(\ref{seesaw-mixing-u}) and (\ref{seesaw-mixing-d}), diagonalize the 
mass matrices in Eq.(\ref{seesaw-mass-part}), and the eigenvalues are identified with $m_{t,b}
(\text{TSS})$ and $m_{T,B} $ where $m_{T,B} \left(> m_{t,b}(\text{TSS}) \right)$ is mass of the 
vector-like partner of the third generation quark. Therefore $c^{t,b}_L,s^{t,b}_R$ should satisfy
\beq
&&
[c^t_L]^2 \equiv \frac{m^2_T - \Sigma^2_U}{m^2_T - m^2_t(\text{TSS})}
\quad , \quad
[s^t_R]^2 \equiv \frac{m^2_t(\text{TSS})}{\Sigma^2_U} [c^t_L]^2
\,,\label{mixing-top}
\\[1ex]
&&
[c^b_L]^2 \equiv \frac{m^2_B - \Sigma^2_D}{m^2_B - m^2_b(\text{TSS})}
\quad , \quad
[s^b_R]^2 \equiv \frac{m^2_b(\text{TSS})}{\Sigma^2_D} [c^b_L]^2
\,.\label{mixing-bottom}
\eeq
Thus the fermion mixing angles are determined by the solutions to the RGEs,  compositeness 
conditions and on-shell conditions in Eqs.(\ref{mixing-top}) and (\ref{mixing-bottom}) for arbitrary 
values of $m_{T,B}$. In Fig.\ref{fermion-mixing-angle}, we show the resultant 
$c^t_L,s^t_R,c^b_L,s^b_R$ for the benchmark parameters  given in 
Eqs.(\ref{tptb-sample0})%
-(\ref{tptb-sample3}).  
Within each $(\tan \phi\,,\,\tan \beta)$-group we consider values $m_{T,B} = 0.8,1,2,5 \TeV$.
The horizontal dot-dashed line in (c-1,2) is the $95 \% \cl $ allowed line by the constraint for 
$\delta g^b_L = (1/2) (s^b_L)^2$. This correction to $g^b_L$ arises at tree level in the present model 
(for details, see section \ref{section-Zbb}), and the allowed region is above this line. %
%
%
In order that fermion sector does not generate a large contribution to the $T$-parameter, we set 
\beq
m_T = m_B = 5 \TeV 
\,. \label{mTB-sample}
\eeq
Since Eq.(\ref{mixing-bottom}) implies $[c^b_L]^2 \simeq 1 -(\Sigma_D/m_B)^2$, i.e. $[s^b_L]^2 \simeq (\Sigma_D/m_B)^2$ for $m_B \gg m_b({\rm TSS})$, 
we see that the above choice for $m_B$ is also not affected by the $\delta g^b_L$ constraint;
see Fig.\ref{fermion-mixing-angle}(c-1,2). 

%

%
%
%

\begin{figure}[htbp]
\begin{center}
\begin{tabular}{cc}
{
\begin{minipage}[t]{0.5\textwidth}
\begin{flushleft} (a) \end{flushleft} \vspace*{-5ex}
\includegraphics[scale=0.65]{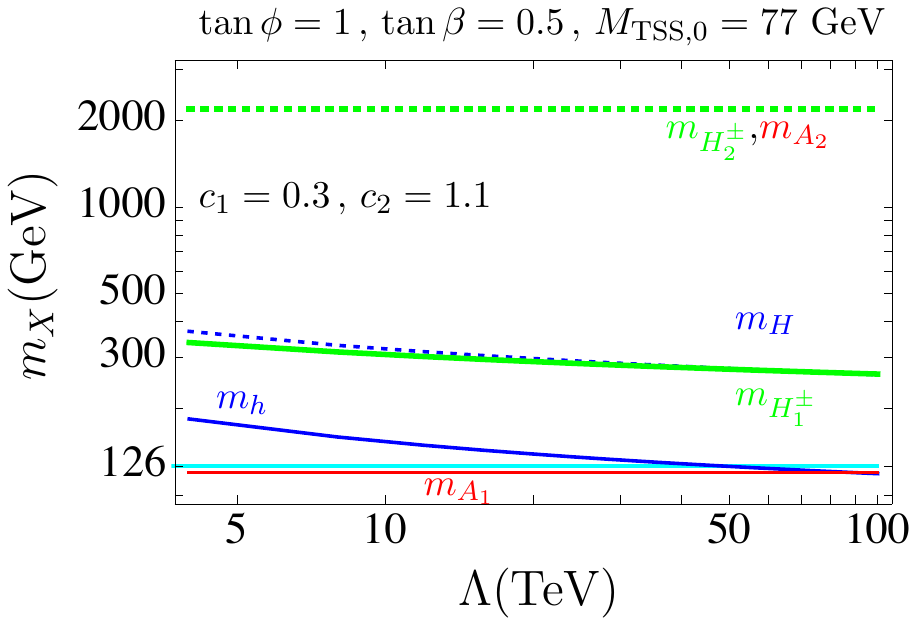} 
\vspace*{2ex}
\end{minipage}
}
{
\begin{minipage}[t]{0.5\textwidth}
\begin{flushleft} (b) \end{flushleft} \vspace*{-5ex}
\includegraphics[scale=0.65]{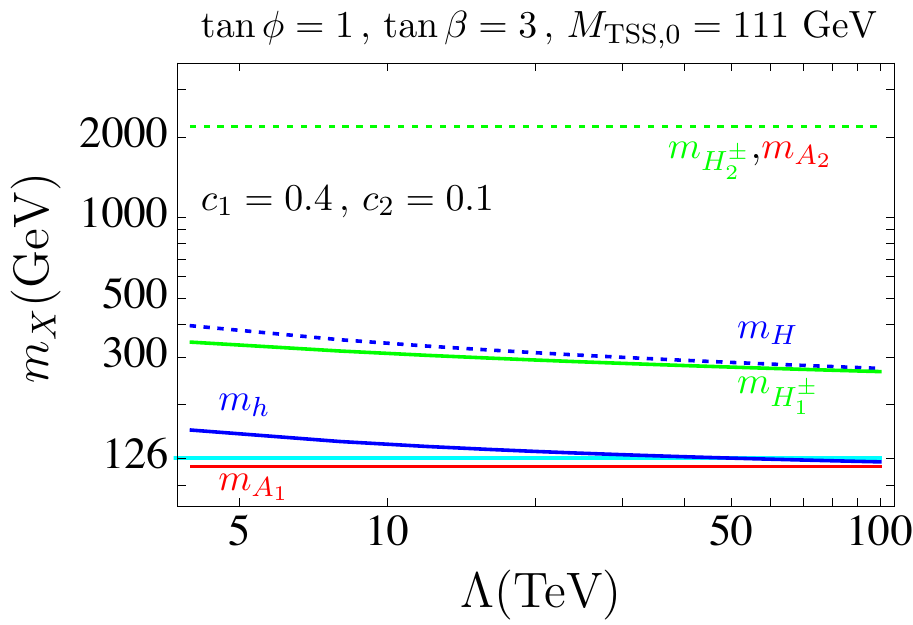} 
\vspace*{2ex}
\end{minipage}
}
\\{
\begin{minipage}[t]{0.5\textwidth}
\begin{flushleft} (c) \end{flushleft} \vspace*{-5ex}
\includegraphics[scale=0.65]{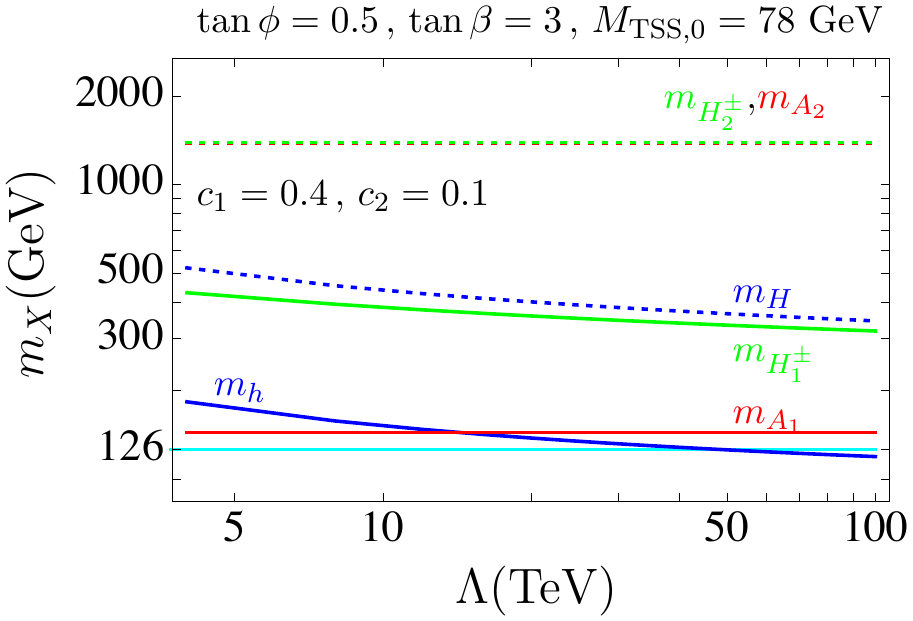} 
\vspace*{2ex}
\end{minipage}
}
{
\begin{minipage}[t]{0.5\textwidth}
\begin{flushleft} (d) \end{flushleft} \vspace*{-5ex}
\includegraphics[scale=0.65]{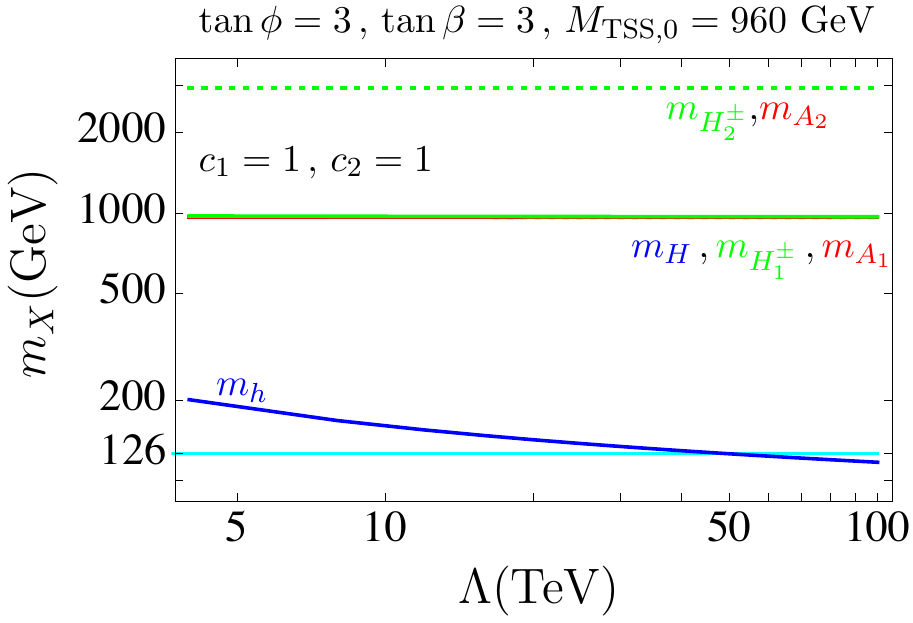} 
\vspace*{2ex}
\end{minipage}
}
\end{tabular}
\caption[]{
Higgs boson mass in the top-seesaw assisted walking TC model for $4 \TeV \leq \Lambda \leq 100 \TeV$. Each panel corresponds to $(\tan \phi,\tan\beta,c_1, c_2) =$ (a) $(1,0.5,2,0.1)$, (b) $(1,3,0.1,0.1)$, (c) $(0.5,3,0.1,0.1)$ and (d) $(3,3,1,1)$. $m_{h,H,A_1,A_2,H^\pm_1,H^\pm_2}$ are blue solid, blue dotted, red solid, red dotted, orange solid, orange dotted curves, respectively.The horizontal cyan solid line shows $m_h = 126 \GeV$ in all panels.
\label{all-higgs-mass}}
\end{center}
\end{figure}

%
To finish this section, in Fig.\ref{all-higgs-mass} we show the higgs boson mass corresponding to different points in the parameter space of the model: (a) Eq.(\ref{MTSS0-sample0}),  (b) Eq.(\ref{MTSS0-sample1}), (c) Eq.(\ref{MTSS0-sample2}) and (d) Eq.(\ref{MTSS0-sample3}). The mass of the vectorlike $B$ quarks is given in Eq. (\ref{mTB-sample}). Of course these results depend on values of $(c_1,c_2)$, and we consider values $(c_1,c_2)$ which minimize the $T$-parameter as will be shown in the section \ref{section-ST}. For the case (d), the dependence of the results on $(c_1,c_2)$ is small. In Fig. \ref{all-higgs-mass}, the blue solid curves correspond to $m_h$,  blue dotted curves to $m_H$, red solid curves to $m_{A_1}$, red  dotted curves to $m_{A_2}$,  green solid curves to $m_{H^\pm_1}$ and green dotted curves to $m_{H^\pm_2}$. In all panels, $m_{A_2}$ and $m_{H^\pm_2}$ have almost degenerate mass around $m^2_{\rm ETC} = \Lambda^2_{\rm TC} = 4 \pi v^2_{\rm TC}$ of Eq. (\ref{TC-contribution-PNGB}). In the case of parameter values in (d), corresponding to $(\tan \phi , \tan \beta) = (3,3)$, we find hierarchical structure of the higgs boson masses as $m_h \simeq 126 \GeV < m_H \simeq m_{A_1} \simeq m_{H^\pm_1} < m_{A_2},m_{H^\pm_2}$. On the other hand, in the case of (a), (b) and (c) parameter values, corresponding to $(\tan \phi , \tan \beta) =(1,0.5),  (1,3), (0.5,3)$, we find the hierarchical structure $m_h,m_{A_1} \simeq 126 \GeV < m_H \simeq m_{H^\pm_1} < m_{A_2},m_{H^\pm_2}$. %
Now, focus on the case of Eq.(\ref{MTSS0-sample4}), i.e. $m_H = 126 \GeV$ at $\Lambda = 50 \TeV$. 
In this case the charged higgs boson is light, with a mass $m_{H^\pm_1} \simeq m_H = 126 \GeV$, 
i.e. $m_{H^\pm_1} < m_t $. In the present model this charged higgs is analogous with the charged top 
pion in the topcolor model. From \cite{Chivukula:2012cp}, the light charged higgs with $m_{H^\pm} 
\simeq 130 \GeV$ is ruled out for $\sin \omega \simeq 0.3$ where the parameter
$\sin \omega$  of \cite{Chivukula:2012cp} corresponds to $v_2/v_{\rm EW}$ in the present model. For $(\tan \phi, \tan \beta) = (3,3)$, we have $v_2/v_{\rm EW} = 0.3$, 
and we conclude that $m_{H^\pm_1} \simeq 130 \GeV$ is ruled out by the charged higgs boson 
search and thus we eliminate the representative point, Eq.(\ref{MTSS0-sample4}), and will not consider it further in this paper.

%
%
%
%

%
\section{EWPT and $\delta g^b_L$ constraints }
\label{EWPT-gLb}
%

In this section we shall constrain the representative points, Eqs.(\ref{MTSS0-sample0})-
(\ref{MTSS0-sample3}) from the electroweak precision
tests (EWPT) and $\delta g^b_L$ including the one-loop corrections.

%
\subsection{EWPT parameter for the higgs sector in the present modell}
\label{section-ST}
%

\begin{figure}[htbp]
\begin{center}
\begin{tabular}{ccc}
{
\begin{minipage}[t]{0.3\textwidth}
\begin{flushleft} (a) \end{flushleft} \vspace*{-3ex}
\includegraphics[scale=0.5]{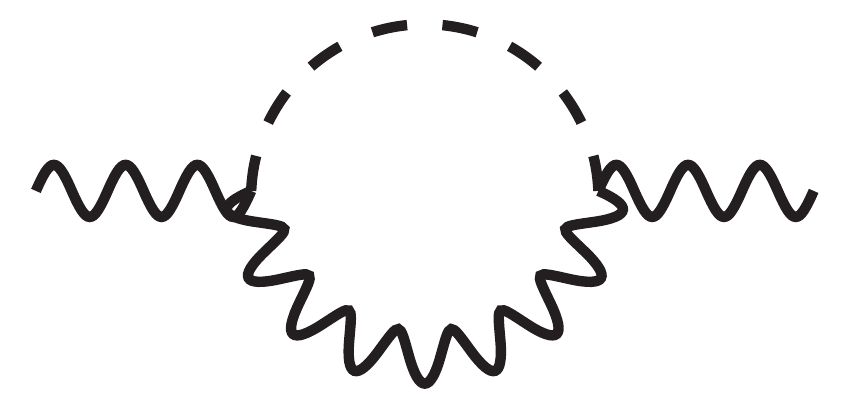} 
\end{minipage}
}
{
\begin{minipage}[t]{0.3\textwidth}
\begin{flushleft} (b) \end{flushleft} \vspace*{-3ex}
\includegraphics[scale=0.5]{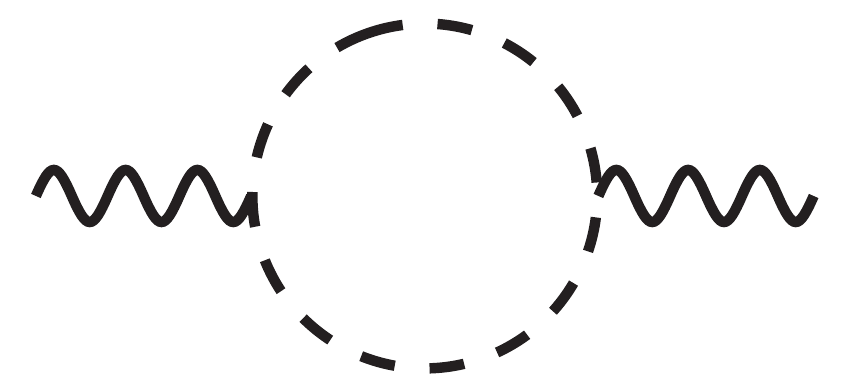} 
\vspace*{2ex}
\end{minipage}
}
{
\begin{minipage}[t]{0.3\textwidth}
\begin{flushleft} (c) \end{flushleft} \vspace*{-3ex}
\includegraphics[scale=0.5]{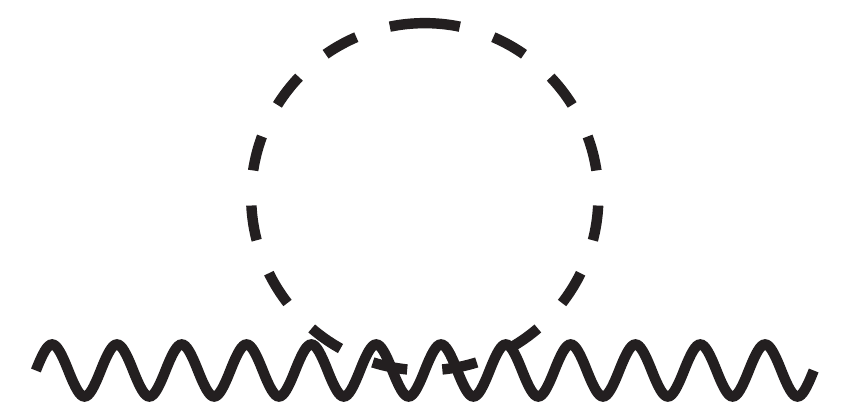} 
\vspace*{2ex}
\end{minipage}
}
\end{tabular}
\caption[]{
The higgs/NGBs contribution to the vacuum polarization. The dashed line and wave line correspond to higgs/NGBs and gauge boson, respectively.
\label{higgs-VP}}
\end{center}
\end{figure}

In this section, we consider the EWPT constraints in the present model. By using Eq.(\ref{hybrid-full-EFT}) in the mass basis of PNGBs and higgses, we obtain the Feynman rules in Tables. \ref{FR-SVV},\ref{FR-SSV} and \ref{FR-SSVV} in appendix \ref{ewptappendix}. Then we compute higgs contributions to the vacuum polarization at the one loop as shown in Fig.\ref{higgs-VP}.
The results for the Peskin-Takeuchi $S$ and $T$ parameters \cite{Peskin:1990zt} are given in 
appendix \ref{ewptappendix}. The results are similar to the generic three higgs doublet model. 
The difference arises from the fact that since we treat the TC sector using the non-linear
sigma model, one of the higgs doublets does not contain the CP-even higgs boson; see 
Eq.(\ref{TC-higgs}). This leads to non-cancelling $1/\epsilon$-contribution in the $S$ and
$T$ parameters. However, this is not a problem, but merely reflects that our effective model is not
ultraviolet complete theory, but should be only studied below a finite cutoff scale.
Therefore, to interpret the final results in terms of the cutoff of effective theory, we replace these divergent part as
\beq
\frac{1}{\bar{\epsilon}} + 1 \to \ln \Lambda^2_{\rm TC}\,,
\eeq
where $\Lambda_{\rm TC}$ is the cutoff of the effective theory for the TC sector and we take $\Lambda_{\rm TC} = 4\pi v_{\rm TC}$. 

\begin{figure}[htbp]
\begin{center}
\includegraphics[scale=0.75]{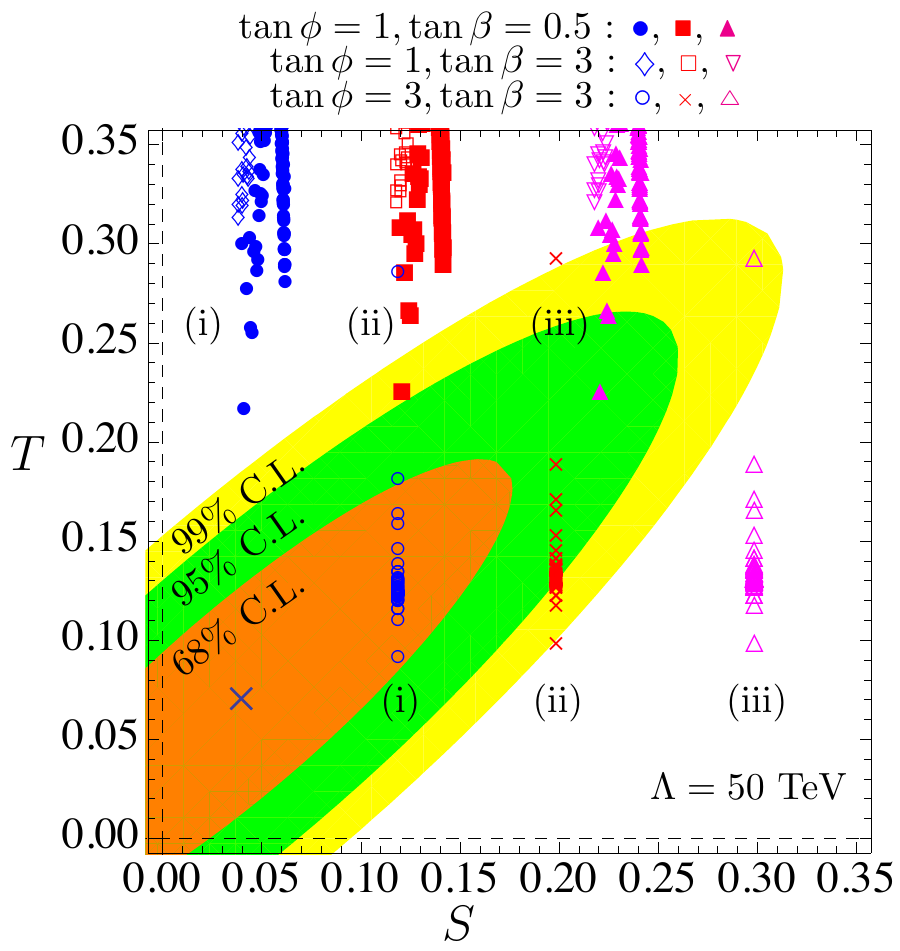} 
\caption[]{
The EWPT constraint for $(\tan \phi\,, \tan \beta ) = (1,0.5), (1,3), (3,3)$ with $m_T = m_B = 5 \TeV$ and $\Lambda = 50 \TeV$. A case of $(\tan \phi \,, \tan \beta) =(0.5,3)$ is on $T \geq 0.6$. We vary $c_1,c_2$ in a range $[0.1,5]$. Groups (i),(ii),(iii) correspond to $S_{\rm TC} = 0$, $S_{\rm TC} = 0$ with $(m_N,m_E) = (120 \GeV, 100 \GeV)$, $S_{\rm TC} = 0.1$ with $(m_N,m_E) = (120 \GeV, 100 \GeV)$, respectively. The shaded region corresponds to $68, 95, 99 \% \cl$ allowed region from inner to outer. $\times$ shows $S= 0.04\,,T = 0.07$. 
\label{ST-contour}}
\end{center}
\end{figure}

In Fig.\ref{ST-contour} we show the EWPT constraint for the present model with the representative points of the parameters as given in Eqs.(\ref{MTSS0-sample0}), (\ref{MTSS0-sample1}) and (\ref{MTSS0-sample3}) with $m_T = m_B = 5 \TeV$, $\Lambda = 50 \TeV$ and varying $(c_1,c_2)$ in the range $0.1 \leq c_{1,2} \leq 5$. We focus on this range, since for $c_{1,2} > 5$, the values of $T$ become large, around $T > 0.4$. 
The shaded regions corresponds to $68, 95, 99 \% \cl$ allowed region from inner to outer ellipses, and experimental results of $S,T$ are \cite{PDG2012}
\beq
S= 0.04 \pm 0.09 
\quad , \quad 
T = 0.07 \pm 0.08\,,
\eeq
and these central values are presented by the cross in Fig.\ref{ST-contour}. We take the reference higgs boson mass as $m^{\rm ref}_h = 117 \GeV$. The results are insensitive if this reference value is varied in a range $115.5 \GeV < m^{\rm ref}_h  < 127 \GeV$ as in \cite{PDG2012}. 
So far our discussion of the technicolor sector has been general. For illustration, 
here we also consider how the
more detailed features may affect the results. As an example, consider minimal walking technicolor,
where a fourth chiral generation of leptons arises due to cancellation of a global anomaly.
Hence, in the present model,  $S$ and $T$ are given by
\beq
S 
\!\!&=&\!\! 
S_{\text{TC}}+S_{\text{3HDM}} +S_{N,E}
+ S_{q4}
+ S_{G',Z'}
+ \Delta_S 
\,, \\[1ex]
T 
\!\!&=&\!\! 
T_{\text{TC}}+T_{\text{3HDM}} +S_{N,E}
+ T_{q4}+ T_{G',Z'}
+ \Delta_T
\,,
\eeq
The factors with subscript TC correspond to the contribution from the TC sector and will be discussed below. The factors with subscript 3HDM correspond to the contributions from the three Higgs doublet sector, and are given in Eqs.(\ref{S-full}) and (\ref{T-full}). The factors with subscript $N,E$ and $q4$ and $G',Z'$ correspond to the contribution from the fourth generation chiral leptons, vectorlike quarks and 
heavy topcolor gauge bosons, respectively, and these are explicitly given in \cite{Fukano:2012qx}.
Note that the contribution from new chiral leptons arises only if we associate the technicolor sector with 
minimal walking technicolor. The contributions $S_{G',Z'}$ and $T_{G',Z'}$ become large below $\Lambda \simeq 10 \TeV$ but are small and negligible for $\Lambda \geq 50 \TeV$. Since we concentrate on $\Lambda \simeq 50 \TeV$, we do not consider these contributions. Finally, the factors $\Delta_{S,T}$  contain the contributions from the SM-like CP-even scalar $h^0$, and the subtraction of the SM higgs contribution; see \cite{Fukano:2012qx}.

We assume that the TC sector conserves custodial symmetry, and hence we take $T_{\text{TC}}=0$. For the 
TC contributions to the $S$-parameter,  we consider the generic TC sector without extra leptons, i.e.
(i) $S_{\rm TC} = 0$, $S_{E,N}=T_{E,N}=0$, and TC sector of minimal walking technicolor (ii)
$S_{\rm TC} = 0$ with $(m_N,m_E) = (120 \GeV, 100 \GeV)$, (iii) $S_{\rm TC} = 0.1$ with 
$(m_N,m_E) = (120 \GeV, 100 \GeV)$. These correspond to groups (i),(ii),(iii) in  Fig.\ref{ST-contour}, respectively. %
From Fig.\ref{ST-contour}, we find that the EWPT constraint allow $(\tan \phi \,,\tan\beta) = (1,0.5) , (3,3)$ among the present representative values in Eqs.(\ref{MTSS0-sample0},\ref{MTSS0-sample1},\ref{MTSS0-sample2},\ref{MTSS0-sample3}). In the case of $(\tan \phi \,,\tan\beta) = (1,3) , (0.5,3)$, the minimum values of $T$ are given $T \simeq 0.31, 0.6$, respectively. The origin of these rather large values can be 
traced to the spectrum: From Fig.\ref{all-higgs-mass} (b) and (c), corresponding to $(\tan \phi \,,\tan \beta) =(1,3), (0.5,3)$, we find $m_{A_1} < m_{H,H^\pm_1}$ and this splitting causes a large contribution to $T$-parameter similarly with the top-seesaw model \cite{He:2001fz}.
 From Fig.\ref{all-higgs-mass} (a), corresponding to $(\tan \phi \,,\tan \beta) =(1,0.5)$, we also find $m_{A_1} < m_{H,H^\pm_1}$ but in this case $m_{A_1} \simeq m_h$ at around $\Lambda = 50 \TeV$, so in this case the overall contribution to $T$-parameter remains smaller, and the result can remain within the $S-T$ ellipsis in Fig.\ref{ST-contour}.
The above results are not affected by variation of $\epsilon_b $, since the fermion contribution to $S,T$-
parameters do not depend on $s^b_R$ and the dependence of $c^b_L$  on $\epsilon_b$ is negligibly 
small. 

%
\subsection{$Z\bar{b}_Lb_L$ constraint}
\label{section-Zbb}
%
Generally, light charged higgs bosons with mass around $300 \GeV$ are constrained  by the experimental value of $R_b$ and $A_b$; for the case of 2HDM, see \cite{Grant:1994ak}. In this section, we discuss the radiative correction to $\delta g^b_L$, which is defined as
\beq
\frac{g}{c_W} Z_\mu \bar{b}_L [g^b_L + \delta g^b_L] b_L
\,,
\eeq
for the higgs sector in the present model. Now, the interactions between fermions and electroweak gauge bosons in the fermion mass basis are given by 
\beq
{\cal L}_{Vff}
\!\!&=&\!\!
\frac{2}{3} e A_\mu \left[ \bar{t} \gamma^\mu t + \bar{T} \gamma^\mu T \right]
-
\frac{1}{3} e A_\mu \left[ \bar{b} \gamma^\mu b + \bar{B} \gamma^\mu B \right]
+
\left[ \text{$Z\bar{f}f + W \bar{f}f$ terms}\right]
\,,
\eeq 
where $\left[ \text{$Z\bar{f}f + W \bar{f}f$ terms}\right]$ are given in Table \ref{Vff-couplings} in appendix \ref{ewptappendix} and $g^{t,b}_{L,R}$ is given by
\beq
&&
g^t_L = \frac{1}{2} - \frac{2}{3} s^2_W
\quad , \quad
g^t_R = -\frac{2}{3} s^2_W
\,,\\
 &&
g^b_L = -\frac{1}{2} + \frac{1}{3} s^2_W
\quad , \quad
g^b_R = \frac{1}{3} s^2_W
\,.
\eeq
For our analysis, we also need the yukawa interactions among 2+1 higgs doublets and fermions. In the present model,  the yukawa terms for third generation quarks and their vector-like partners, which is a part of Eq.(\ref{reno-yukawa-H4G}), are 
\beq
{\cal L}^{\rm 3-4}_{\rm yukawa}
=
- y_1 \bar{Q}^{(3)}_L \Phi_1 D^{(4)}_R - y_2 \bar{Q}^{(3)}_L \tilde{\Phi}_2 U^{(4)}_R
- y^b_{\rm TC} \bar{q}_L \Phi_{\rm TC} b_R - y^t_{\rm TC} \bar{q}_L \tilde{\Phi}_{\rm TC} t_R
+ \text{h.c.}
\,,\label{3higgs-yukawa}
\eeq 
where the first and second terms are written in the topcolor interaction basis for fermions but the third and fourth terms are in the mass basis for fermions. 
The couplings $y_{1,2}$ are solved from RGEs, Eqs.(\ref{RGE-QCD})-
(\ref{RGE-lambda4}) under the compositeness conditions, Eqs.(\ref{cc-yukawa})-
(\ref{cc-lambda34}) with on-shell condition Eqs.(\ref{on-shell-condition-tb}). On the other hand, the couplings $y^{b,t}_{\rm TC}$ are given by
\beq
y^b_{\rm TC} = \frac{\sqrt{2} \epsilon_b m_b}{v_{\rm TC}}
\quad , \quad
y^t_{\rm TC} = \frac{\sqrt{2} \epsilon_t m_t}{v_{\rm TC}}
\,,
\eeq
where $\epsilon_{t,b}$ are defined in Eqs. (\ref{physical-top-mass}) and (\ref{physical-bottom-mass}). For our purpose, it is enough to consider yukawa interactions which include the charged scalar particles and  the left-handed bottom quark, and these yukawa interactions are given in Table \ref{charged-Gff-couplings} in appendix \ref{ewptappendix}.
The experimental $95\% \cl$ constraint for $[\delta g^b_L]$ by both $R_b$ and $A_b$ is given by \cite{Dawson:2012di}
\beq
-2.7 \times 10^{-3} \leq \delta g^b_L \leq 1.4 \times 10^{-3}
\quad (\text{95\%\!\!\cl})
\label{gLb-95CLconstraint}
\,.
\eeq 
Throughout the calculation in this section, we will work under the assumption $m^2_b =0$. 

In the present model, one can see easily from Table \ref{Vff-couplings} that $\delta g^b_L$ at the tree level is given by
\beq
[\delta g^b_L]_{\text{tree}} = \frac{1}{2} (s^b_L)^2
\label{deltagLb-tree}
\,.
\eeq
As to the one-loop radiative correction, we divide it into two parts: one including the EW gauge boson in the loop, and another that does not include any EW gauge bosons. We denote these two corrections as $[\delta g^b_L]^{\text{1loop}}_{\text{gauge}}$ and $[\delta g^b_L]^{\text{1loop}}_{\text{NGB}}$, respectively. Diagrammatically, $[\delta g^b_L]^{\text{1loop}}_{\text{gauge}}$ is  
\beq
[\delta g^b_L]^{\text{1loop}}_{\text{gauge}} 
\!\!&=&\!\!
\parbox[c]{15ex}{\includegraphics[width=15ex]{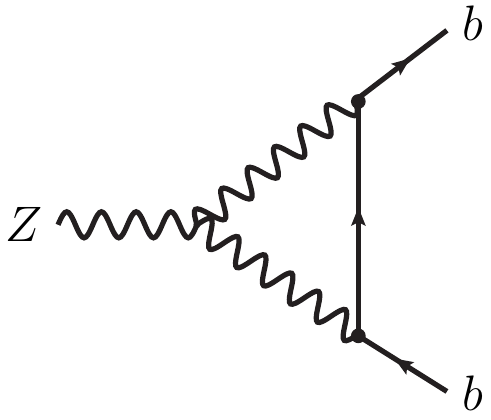}}
\hspace*{1ex} +\hspace*{1ex}
\parbox[c]{15ex}{\includegraphics[width=15ex]{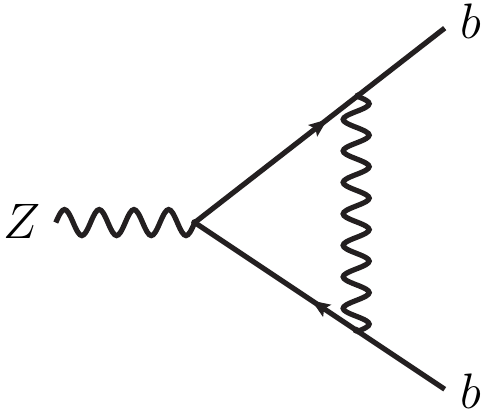}}
\hspace*{1ex} +\hspace*{1ex}
\parbox[l]{15ex}{\includegraphics[width=15ex]{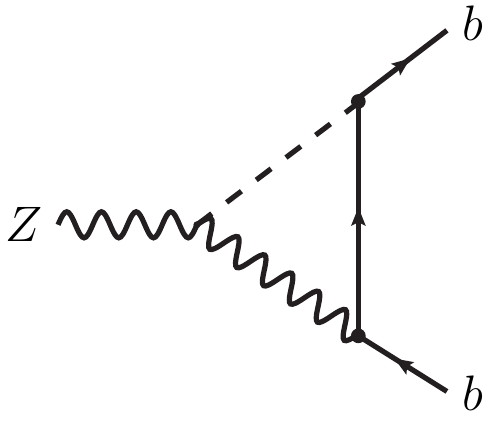}}
\hspace*{1ex} +\hspace*{1ex}
\parbox[r]{15ex}{\includegraphics[width=15ex]{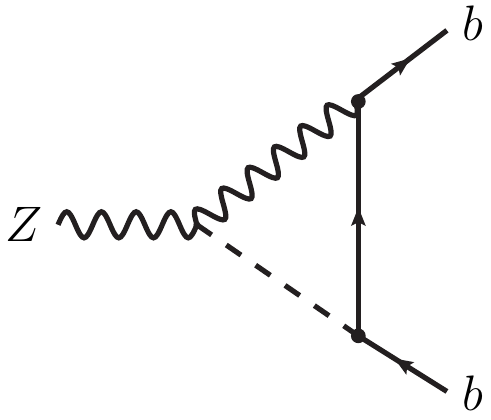}}
\,\nonumber\\[1ex]
&&
\hspace*{1ex} +\hspace*{1ex}
\parbox[l]{15ex}{\includegraphics[width=15ex]{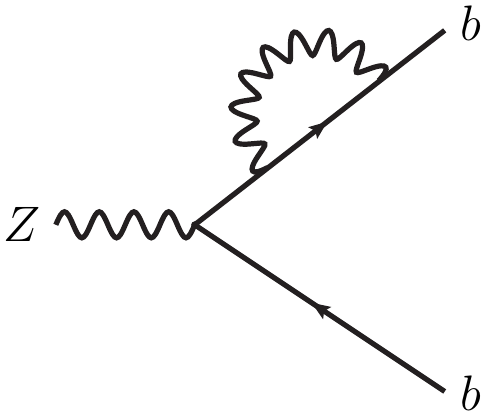}}
\hspace*{1ex} +\hspace*{1ex}
\parbox[r]{15ex}{\includegraphics[width=15ex]{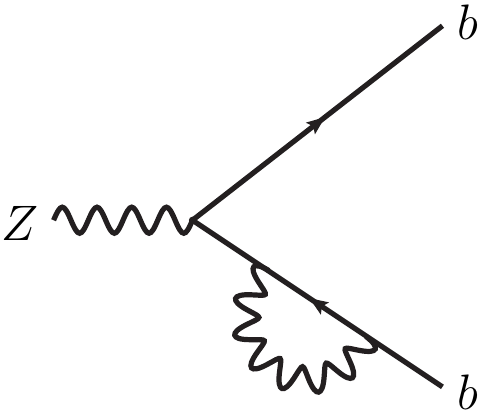}}
\quad ,\label{Zbb-1loop-top-only}
\eeq
and $[\delta g^b_L]^{\text{1loop}}_{\text{NGB}}$ is 
\beq
[\delta g^b_L]^{\text{1loop}}_{\text{NGB}} 
\!\!&=&\!\!
\parbox[c]{15ex}{\includegraphics[width=15ex]{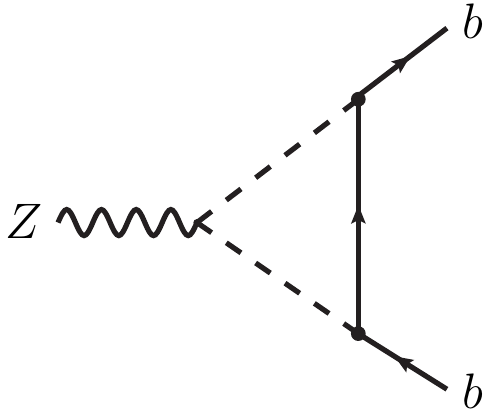}}
\hspace*{1ex} +\hspace*{1ex}
\parbox[c]{15ex}{\includegraphics[width=15ex]{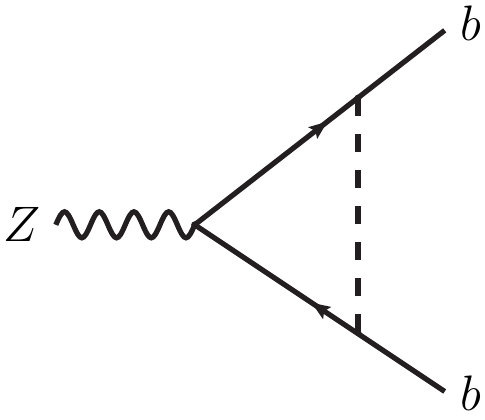}}
\hspace*{1ex} +\hspace*{1ex}
\parbox[l]{15ex}{\includegraphics[width=15ex]{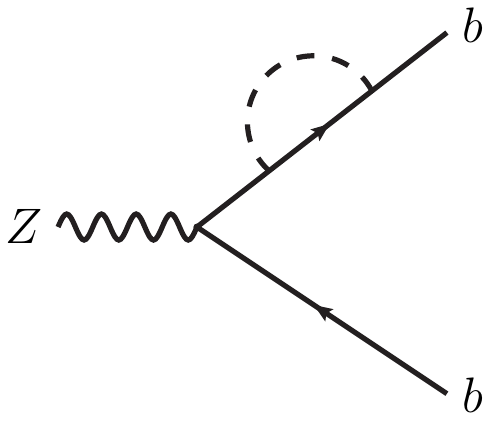}}
\hspace*{1ex} +\hspace*{1ex}
\parbox[r]{15ex}{\includegraphics[width=15ex]{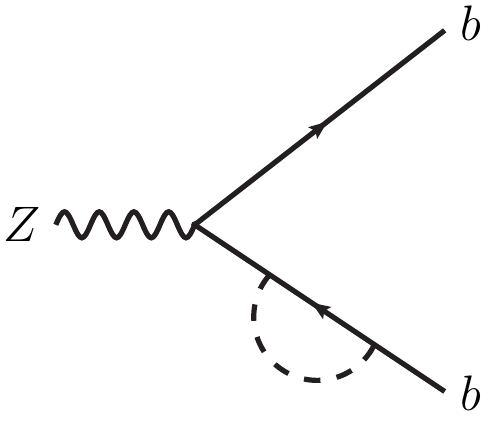}}
\quad .\label{Zbb-1loop-top-and-T}
\eeq
We take the incoming $Z$ boson momentum equal to zero.
Using the results from Appendix \ref{ewptappendix}, we obtain the renormalized 
$[\delta g^b_L]^{\text{1loop}}$ as follows: First, the UV-divergences are renormalized 
as \cite{Bernabeu:1987me}
\beq
[\delta g^b_L]^{\rm 1loop}_{\rm reno.}
\equiv 
[\delta g^b_L]^{\rm 1loop}
-
[\delta g^b_L]^{\rm 1loop}_{m^2_t , m^2_T = 0}
\,.\label{deltagLb-1loop}
\eeq
Thus the deviation from $g^b_L$ in the SM in the present model is given by
\beq
[\delta g^b_L]
=
[\delta g^b_L]_{\rm tree}
+
\Delta [\delta g^b_L]^{\rm 1loop}.
\label{deltagLb-tree-1loop}
\eeq
Here $\Delta [\delta g^b_L]^{\rm 1loop}$ is defined as
\beq
\Delta [\delta g^b_L]^{\rm 1loop}
\equiv
[\delta g^b_L]^{\rm 1loop}_{\rm reno.}
-
[\delta g^b_L]^{\rm SM-1loop}_{\rm reno.}\,,
\eeq
and $[\delta g^b_L]^{\rm SM-1loop}_{\rm reno.}$ is given by $[\delta g^b_L]^{\rm 1loop}_{\rm reno.}$ in the limit $\cos \theta^{t,b}_L \to 1\,,\Sigma_U \to 0\,, \cos \phi \to 0\,,\epsilon_t \to 1\,,v_{\rm TC} \to v_{\rm EW}$. 
Let us perform a nontrivial check on our results by taking a limit $c^b_L \to 1$, $s^{t,b}_R \to 1$, $\tan \zeta \to \infty$ and $\tan \beta \to \infty$ with $1-\epsilon_t \gg \epsilon_t$. This  corresponds to the well-known TC2 model \cite{Hill:1994hp}. In this limit, $[\delta g^b_L]_{\rm gauge}$ in Eq.(\ref{Zbb-1loop-gauge-result}) reduces to the SM one-loop result, but $[\delta g^b_L]_{\rm NGB}$ in Eq.(\ref{Zbb-1loop-NGB-only-result}) contains also a contribution beyond the Standard Model. Thus under this limit, we find a result
\beq
[\delta g^b_L]_{\rm TC2}
\!\!&=&\!\!
- \frac{1}{2}\frac{1}{16 \pi^2}  [Y^G_{tb}]^2  C_{01}(m^2_t,M^2_W) 
\,,\nonumber\\[1ex]
\!\!&=&\!\!
\frac{1}{2}\frac{1}{16 \pi^2}  \left[ \frac{\sqrt{2}m_t}{v_2} \frac{v_{\rm TC}}{v_{\rm EW}}\right]^2  
\left( 
- \frac{x}{(x-1)^2}\ln x + \frac{x}{x - 1}
\right)
\,,
\eeq 
where $x\equiv m^2_t/M^2_{H^\pm_2}$, and this result reproduces the result obtained in 
\cite{Loinaz:1998jg}. 

\begin{figure}[htb]
\begin{center}
\includegraphics[scale=0.8]{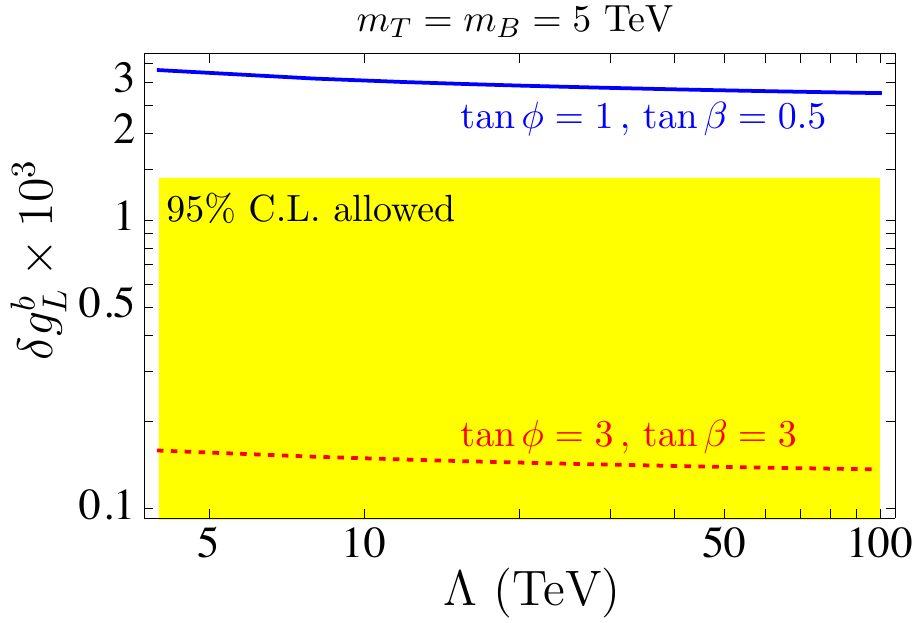} 
\vspace*{2ex}
\caption[]{
$\delta g^b_L$ constraint for the present model for $(\tan \phi\,,\tan \beta) = $ (solid) $(1,0.5)$ and  (dotted) $(3,3)$ with $m_T = m_B = 5 \TeV$.  The shaded region shows the $95 \% \cl$ allowed region in accordance with Eq.(\ref{gLb-95CLconstraint}). 
\label{gLb-constraint}}
\end{center}
\end{figure}

Turning to our model study, then, in Fig.\ref{gLb-constraint}, we show constraint for $[\delta g^b_L]$ defined as Eq.(\ref{deltagLb-tree-1loop}) with parameter values from  Eqs.(\ref{MTSS0-sample0})and (\ref{MTSS0-sample3}), which are allowed by the EWPT constraint as seen from Fig.\ref{ST-contour}. The shaded region shows the $95 \% \cl$ allowed region in accordance with 
Eq.(\ref{gLb-95CLconstraint}). In Table \ref{summary-constraint}, we summarize the judgement of the  experimental constraints we have considered for the representative parameter values from
 Eqs.(\ref{tptb-sample0})-
 (\ref{tptb-sample3}) and  Eqs.(\ref{MTSS0-sample0}-
 (\ref{MTSS0-sample3}). Thus, among the representative values Eqs.(\ref{MTSS0-sample0})-
 (\ref{MTSS0-sample3}) the EWPT and $\delta g^b_L$ constraints favor only a case of
\beq
\tan \phi = 3 \quad , \quad \tan \beta =3 \quad \text{with} \quad M_{{\rm TSS},0} = 960 \GeV 
\,,
\eeq
which derives $m_h = 126 \GeV$ at $\Lambda = 50 \TeV$, and this light CP-even higgs boson arises mainly from $\vev{\bar{U}^{(3)}_L U^{(4)}_R} \neq 0$ since $|\tan \alpha| < 1$ as shown in Fig.\ref{CP-even-higgs}(b). The results are insensitive to variations of $\epsilon_b $ since $[\delta g^b_L]$ does not depend on $s^b_R$, and the dependence of  $c^b_L$ on $\epsilon_b$ is negligibly small.

\begin{table}[hptb]
\begin{tabular}{cc}
{
\tabcolsep=1ex
\renewcommand\arraystretch{1.2}
\begin{tabular}{|c|c|c||c|c|c|c|}
\hline 
$\tan \phi$ & $\tan \beta$ & $M_{{\rm TSS},0}$ & at $\Lambda = 50 \TeV$ & $m_{H^\pm_1} > m_t$ & EWPT (Fig.\ref{ST-contour}) & $\delta g^b_L$   (Fig.\ref{gLb-constraint})
\\ \hline \hline
$1$ & $0.5$ & $77 \GeV$ & $m_h = 126 \GeV$ & Yes  & Yes (4th leptons are necessary)& No
\\ \hline 
$1$ & $3$ & $111 \GeV$ & $m_h = 126 \GeV$ & Yes & No & -
\\ \hline 
$0.5$ & $3$ & $78 \GeV$ & $m_h = 126 \GeV$ & Yes & No & -
\\ \hline 
$3$ & $3$ & $73 \GeV$ & $m_H = 126 \GeV$ & No & - & -
\\ \hline 
$3$ & $3$ & $960 \GeV$ & $m_h = 126 \GeV$ & Yes & Yes (4th leptons are not necessary) & Yes
\\ \hline 
\end{tabular}
}
\end{tabular}
\caption{
Summary of the representative values of $(\tan \phi\,,\,\tan \beta)$ in Eqs.(\ref{tptb-sample0},\ref{tptb-sample1},\ref{tptb-sample2},\ref{tptb-sample3}). "Yes" means  allowed by the constraint and "No" means not allowed by the constraint. The hyphen ("-") means that this is not needed. 
As to $m_{H^\pm_1}$ constraint, see section \ref{dynamical-results}.
\label{summary-constraint}
}
\end{table}

%
\section{The model and $126 \GeV$ Higgs at the LHC}
\label{126higgsLHC}
%

In this section, we focus on the light CP-even higgs boson $h^0$ in the present model, and compare the present model with the recent LHC higgs search results for the representative values 
\beq
\tan \phi = 3 \quad , \quad \tan \beta =3 \quad \text{with} \quad 
M_{{\rm TSS},0} = 960 \GeV \quad \text{and} \quad m_T = m_B = 5 \TeV
\,.
\eeq
In this section we fix $\epsilon_t = 0.5$ but we vary $\epsilon_b$ in a range $0.1 \leq \epsilon_b \leq 1$ which does not affect the experimental constraints discussed in section \ref{EWPT-gLb}. 
%
For the LHC phenomenology, the relevant part of the full Lagrangian is
\beq
{\cal L}
\!\!&=&\!\!
C_{h WW}\left( gM_W \cdot W^+_\mu W^{-\mu} + \frac{g}{2c_W}M_Z \cdot Z_\mu Z^{\mu}\right) h
\nonumber\\
&&
-C_{h ff} \frac{m_f}{v_{\rm EW}} \cdot h \bar{f}f
- h \left[ 
C^L_{hFf} \bar{F}_R f_L + C^L_{hfF} \bar{f}_R F_L
+ C^R_{hFf} \bar{F}_L f_R + C^R_{hfF} \bar{f}_L F_R
\right]
\nonumber\\
&&
+ C_{Wff}
\frac{g}{\sqrt{2}} \left[ \bar{f}^u_{iL} \gamma^\mu W^+_\mu V_{ij} f^d_{jL}  + \text{h.c.} \right]
\nonumber\\
&&
+
\frac{g}{2 c_W}Z_\mu \bar{f} \gamma^\mu \left[  C^V_{Zff} \hat{v}_f - C^A_{Zff} \hat{a}_f \gamma_5 \right] f
+
\frac{g}{c_W} Z_\mu \sum_{F \neq f} \bar{F} \gamma^\mu 
\left[ C^L_{ZFf} \frac{1-\gamma_5}{2} + C^R_{ZFf} \frac{1+\gamma_5}{2}\right] 
f
\nonumber\\
&&
+ (gM_W)  \sum_{i=1,2} C_{h H^\pm_i H^\pm_i}  \phi H^+_i H^-_i
+  \frac{i g (c^2_W - s^2_W)}{2 c_W }  Z_\mu 
\sum_{i=1,2} \left[(\partial^\mu H^+_i)H^-_i - H^+_i(\partial^\mu H_i^-)\right]
\,,\label{Lag-hdecay}
\eeq
where 
$V_{ij}$ is the CKM (MNS) matrix if $f^{u,d}_i$ are quarks(leptons) and $\hat{v}_f,\hat{a}_f$ are defined as
\beq
\hat{v}_f \!\!&=&\!\! g^f_L + g^f_R = T^f_3 - 2 s^2_W Q_f \,,\, 
\\[1ex]
\hat{a}_f \!\!&=&\!\! g^f_L + g^f_R = T^f_3\,.
\eeq
In the SM case, prefactors $C$ in Eq.(\ref{Lag-hdecay}) are 
\beq
C_{ h WW} = C_{h ff} = C_{Wff} = C^V_{Zff} = C^A_{Zff} = 1
\quad , \quad
\text{others $= 0$}\,.
\eeq
In the present model, on the other hand, the prefactors $C_{hXY} (X,Y \neq H^\pm_i)$ in 
Eq. (\ref{Lag-hdecay}) are given by
\beq
&&
C_{h WW} = \cos \phi \sin (\beta- \alpha) 
\,,\label{chWW}
\\[1ex]
&&
C_{h tt}
=
\frac{y_2 v_{\rm EW}}{\sqrt{2}m_t} c^t_L s^t_R \cos \alpha
\,,\label{chtt}
\\[1ex]
&&
C_{h bb}
=
- \frac{y_1 v_{\rm EW}}{\sqrt{2} m_b} c^b_L s^b_R \sin \alpha
\,,\label{chbb}
\\[1ex]
&&
C_{ hTT}
=
\frac{y_2 v_{\rm EW}}{\sqrt{2}m_T}  s^t_L c^t_R \cos \alpha
\,,\label{chTT}
\\[1ex]
&&
C_{h BB}
=
- \frac{y_1 v_{\rm EW}}{\sqrt{2}m_B}s^b_L c^b_R \sin \alpha
\,,\label{chBB}
\\[1ex]
&&
C^L_{hTt} = C^R_{htT} = \frac{y_2}{\sqrt{2}} c^t_L c^t_R \cos \alpha
\,,\label{chLTt}
\\
&&
 C^L_{htT} = C^R_{hTt} = \frac{y_2}{\sqrt{2}} s^t_L s^t_R \cos \alpha
\,,\label{chLtT}
\\
&&
C^L_{hBb} = C^R_{hbB} = -\frac{y_1}{\sqrt{2}} c^b_L c^b_R \sin \alpha
\,,\label{chLBb}
\\
&&
 C^L_{hbB} = C^R_{hBb} = -\frac{y_1}{\sqrt{2}} s^b_L s^b_R \sin \alpha
\,,\label{chLbB}
\eeq
and $C_{WXY,ZXY}$ in Eq.(\ref{Lag-hdecay}) are read off from Table \ref{Vff-couplings} in the Appendix \ref{ewptappendix}  as 
\beq
&&
C_{Wff}
=
\begin{cases}
c^t_L c^b_L  & \text{for $Wtb$}\\[1ex]
1  & \text{for other light fermions}\\[1ex]
\end{cases}
\,,\label{cWff}
\\[1ex]
&&
C^V_{Z ff}
=
\begin{cases}
1 - (s^t_L)^2/(2\hat{v}_t) & \text{for $f = t$}
\\[1ex]
1 - (c^t_L)^2/(2\hat{v}_t) & \text{for $f = T$}
\\[1ex]
1 + (s^b_L)^2/(2\hat{v}_b) & \text{for $f = b$}
\\[1ex]
1 + (c^b_L)^2/(2\hat{v}_b) & \text{for $f = B$}
\end{cases}
\,,\label{cZffV}
\\[1ex]
&&
C^A_{Z ff}
=
\begin{cases}
(c^t_L)^2/(2\hat{a}_t) & \text{for $f = t$}
\\[1ex]
(s^t_L)^2/(2\hat{a}_t) & \text{for $f = T$}
\\[1ex]
-(c^b_L)^2/(2\hat{a}_t) & \text{for $f = b$}
\\[1ex]
-(s^b_L)^2/(2\hat{a}_t) & \text{for $f = B$}
\end{cases}\,,
\label{cZffA} 
\\[1ex]
&&
C^L_{ZTt} = \frac{1}{2} c^t_L s^t_L
\quad , \quad
C^R_{ZTt} = 0
\,,\label{cLZTt}\\[1ex]
&&
C^L_{ZBb} = -\frac{1}{2} c^b_L s^b_L
\quad , \quad
C^R_{ZBb} = 0
\,.\label{cLZBb}
\eeq
Here we show only couplings which involve the third family quarks and their vector-like partners. For other fermions (leptons,first and second family quarks), the couplings are the same as the SM case. Finally, the coupling term proportional to $C_{h H^\pm_i H^\pm_i}$ in Eq.(\ref{Lag-hdecay}) is derived from the potential $V(\Phi_1,\Phi_2,\Phi_{\rm TC})$ in Eq.(\ref{3HDM-potential}). Its expression is lengthy, and 
we do not write it explicitly.

By using Eqs.(\ref{Lag-hdecay})-(\ref{cLZBb}), we evaluate the decay width of $h^0$. Like the SM-higgs boson, also $h^0$ generally decays into $WW/ZZ,\bar{f}f$ via two body decay, $WW^*/ZZ^*$ via three body decay and $\gamma \gamma, gg, Z\gamma$ via loop processes. The relevant decay widths are collected in the Appendix \ref{phenoappendix}.
%
%
%
%
%
Applying these results,
we now discuss the production cross section and the signal strengths of the lightest higgs boson in the present model. First, we consider the production cross sections. The cross section of gluon fusion process of higgs boson production $\sigma_{\rm ggF}(h)$ is enhanced compared with the SM case as
\beq
r_{\rm ggF} \equiv
\frac{\sigma_{\rm ggF}[\text{TSSTC}]}{\sigma_{\rm ggF}[\text{SM}]}
=
\frac{\Gamma( h \to gg)[\text{TSSTC}]}{\Gamma( h \to gg)[\text{SM}]}
\simeq 
2.3 - 2.8\,,
\quad \text{(for $0.1 \leq \epsilon_b \leq 1$)}
\eeq
since $C_{htt}$ in Eq.(\ref{chtt}) becomes large; $C_{htt} \simeq 2$.
Note that although there are vector-like fermions in the present model, their couplings with the higgs boson, $C_{hTT,hBB}$ in Eq.(\ref{chTT}) and (\ref{chBB}), are very small due to $c^t_L, c^b_L \simeq 1$ as seen from Fig.\ref{fermion-mixing-angle}(a-2),(c-2). This means, that vector-like fermions do not give a large contribution to the loop process $gg \to h, h\to gg/\gamma\gamma/Z\gamma$ in the present model. This result is different from results in a model including vector-like quarks e.g.\cite{Azatov:2012rj}. %
On the other hand, the cross section of vector boson fusion process (VBF) and vector boson associated process (WH/ZH) of higgs boson production are suppressed compared with the SM case as
\beq
r_{\rm VBF} 
\!\!&\equiv&\!\!
\frac{\sigma_{\rm VBF}[\text{TSSTC}]}{\sigma_{\rm VBF}[\text{SM}]}
=
\frac{\Gamma( h \to WW^*/ZZ^*)[\text{TSSTC}]}{\Gamma( h \to WW^*/ZZ^*)[\text{SM}]}
\simeq 0.1\,,
\quad \text{(for $0.1 \leq \epsilon_b \leq 1$)}
\,,
\\[1ex]
 r_{\rm WH/ZH} 
\!\!&\equiv&\!\! 
\frac{\sigma_{\rm WH/ZH}[\text{TSSTC}]}{\sigma_{\rm WH/ZH}[\text{SM}]}
=
\frac{\Gamma( h \to WW^*/ZZ^*)[\text{TSSTC}]}{\Gamma( h \to WW^*/ZZ^*)[\text{SM}]}
\simeq 0.1\,,
\quad \text{(for $0.1 \leq \epsilon_b \leq 1$)}
\,.
\eeq
This suppression arises since these ratios mainly depend on $C_{hWW}$ in Eq.(\ref{chWW}), which shows that $C_{hWW} \propto \cos \phi$, and hence becomes small if $\tan \phi$ becomes large. %
Thus we obtain the ratio of total higgs boson production cross sections for $0.1\leq\epsilon_b\leq 1$ as
\beq
\frac{\sigma [\text{TSSTC}]}{\sigma[\text{SM}]}
\!\!&\equiv&\!\!
\frac{r_{\rm ggF} \cdot \sigma_{\rm ggF} [\text{SM}] + r_{\rm VBF} \cdot \sigma_{\rm VBF}[\text{SM}] + r_{\rm WH}\cdot \sigma_{\rm WH}[\text{SM}] + r_{\rm ZH}\cdot \sigma_{\rm ZH}[\text{SM}]}{\sigma_{\rm ggF} [\text{SM}] + \sigma_{\rm VBF}[\text{SM}] + \sigma_{\rm WH}[\text{SM}]+ \sigma_{\rm ZH}[\text{SM}]}
\simeq 2 -2.5\,,
\eeq
where we have used values of $\sigma[{\rm SM}]$ for $m_h = 126 \GeV$ from \cite{LHC-YellowReport-7TeV}: $\sigma_{\rm ggF}[{\rm SM}] = 15.08 ({\rm pb})$, $\sigma_{\rm VBF}[{\rm SM}] = 1.199 ({\rm pb})$, $\sigma_{\rm WH}[{\rm SM}] = 0.5576 ({\rm pb})$ and $ \sigma_{\rm ZF}[{\rm SM}] = 0.3077 ({\rm pb})$. %

Next, we consider the signal strength $\mu_X$, which is defined as
\beq
\mu_X 
\equiv 
\frac{\sigma[\text{TSSTC}]}{\sigma[\text{SM}]}
\times
\frac{\text{Br} (h \to X)}{\text{Br} (h^{\rm SM} \to X)}
\,,\label{def-signal-strength-for-VV}
\eeq
for $X = \gamma \gamma/WW^*/ZZ^*/\tau^+\tau^-$ and
\beq
\mu_{bb} 
\equiv 
\frac{r_{\rm WH}\cdot \sigma_{\rm WH}[\text{SM}] + r_{\rm ZH}\cdot \sigma_{\rm ZH}[\text{SM}]}{\sigma_{\rm WH}[\text{SM}] + \sigma_{\rm ZH}[\text{SM}]}
\times
\frac{\text{Br} (h \to b\bar{b})}{\text{Br} (h^{\rm SM} \to b\bar{b})}
\,,\label{def-signal-strength-for-bb}
\eeq
for $X = b\bar{b}$. We note that $\mu_{\tau\tau}[\text{TSSTC}] = 0$ in the present model since the higgs boson does not couple to leptons (see Eq.(\ref{reno-yukawa-H4G})). This fact is different from the SM higgs boson case. %
For reference, we list the LHC results of $\mu_X$: 
\beq
\begin{aligned}
\mu_{\gamma \gamma} &= 1.8 \pm 0.5 &&(\text{ATLAS $7 \TeV + 8 \TeV$ \cite{:2012gk}})
\,,\\[1ex]
\mu_{WW^*} &= 1.3 \pm 0.5 &&(\text{ATLAS $7 \TeV + 8 \TeV$ \cite{:2012gk}})
\,,\\[1ex]
\mu_{ZZ^*} &= 1.4 \pm 0.6 &&(\text{ATLAS $7 \TeV + 8 \TeV$ \cite{:2012gk}})
\,,\\[1ex]
\mu_{bb} &= 0.46 \pm 2.18 &&(\text{ATLAS $7 \TeV $ \cite{Espinosa:2012in}})
\,,\\[1ex]
\mu_{\tau\tau} &= 0.45 \pm 1.8 &&(\text{ATLAS $7 \TeV $ \cite{Espinosa:2012in}})
\end{aligned}
\label{signalstrength-ATLAS}
\eeq
for $m_h = 126.5 \GeV$ at the ATLAS group,
\beq
\begin{aligned}
\mu_{\gamma \gamma} &= 1.56 \pm 0.43 &&(\text{CMS $7 \TeV + 8 \TeV$ \cite{CMS-PAS-HIG-12-015}})
\,,\\[1ex]
\mu_{WW^*} &= 0.38 \pm 0.56/0.98 \pm 0.71 &&(\text{CMS $7 \TeV / 8\TeV$ \cite{Espinosa:2012in}})
\,,\\[1ex]
\mu_{ZZ^*} &= 0.7 \pm 0.4 &&(\text{CMS $7 \TeV + 8 \TeV$ \cite{CMS-PAS-HIG-12-016}})
\,,\\[1ex]
\mu_{bb} &= 0.59 \pm 1.17/0.41\pm0.94 &&(\text{CMS $7 \TeV/8 \TeV $ \cite{Espinosa:2012in}})
\,,\\[1ex]
\mu_{\tau\tau} &= 0.62 \pm 1.13/-0.72 \pm 0.97 &&(\text{CMS $7 \TeV /8\TeV$ \cite{Espinosa:2012in}})
\end{aligned}
\label{signalstrength-CMS}
\eeq
for $m_h = 125 \GeV$ at the CMS group. From Eqs.(\ref{signalstrength-ATLAS}) and (\ref{signalstrength-CMS}), we find that $\mu_{\tau \tau}[{\rm TSSTC}] = 0$ is consistent with the present LHC results. %

\begin{figure}[htbp]
\begin{center}
\begin{tabular}{cc}
{
\begin{minipage}[t]{0.5\textwidth}
\begin{flushleft} (a) \end{flushleft} \vspace*{-5ex}
\includegraphics[scale=0.7]{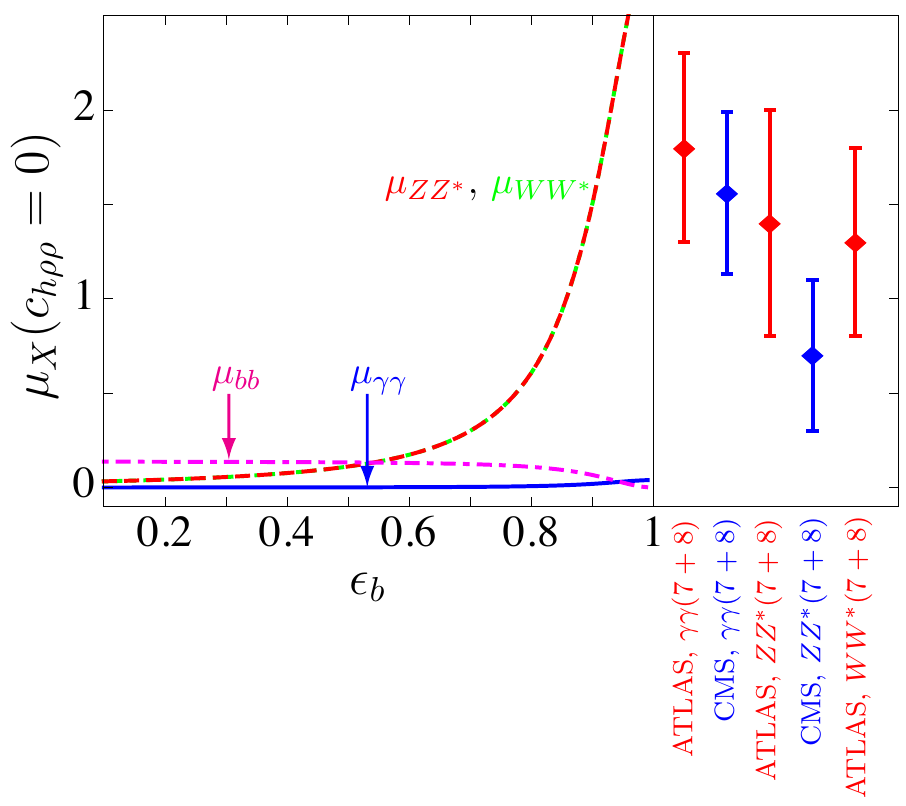} 
\vspace*{2ex}
\end{minipage}
}
{
\begin{minipage}[t]{0.5\textwidth}
\begin{flushleft} (b) \end{flushleft} \vspace*{-5ex}
\includegraphics[scale=0.7]{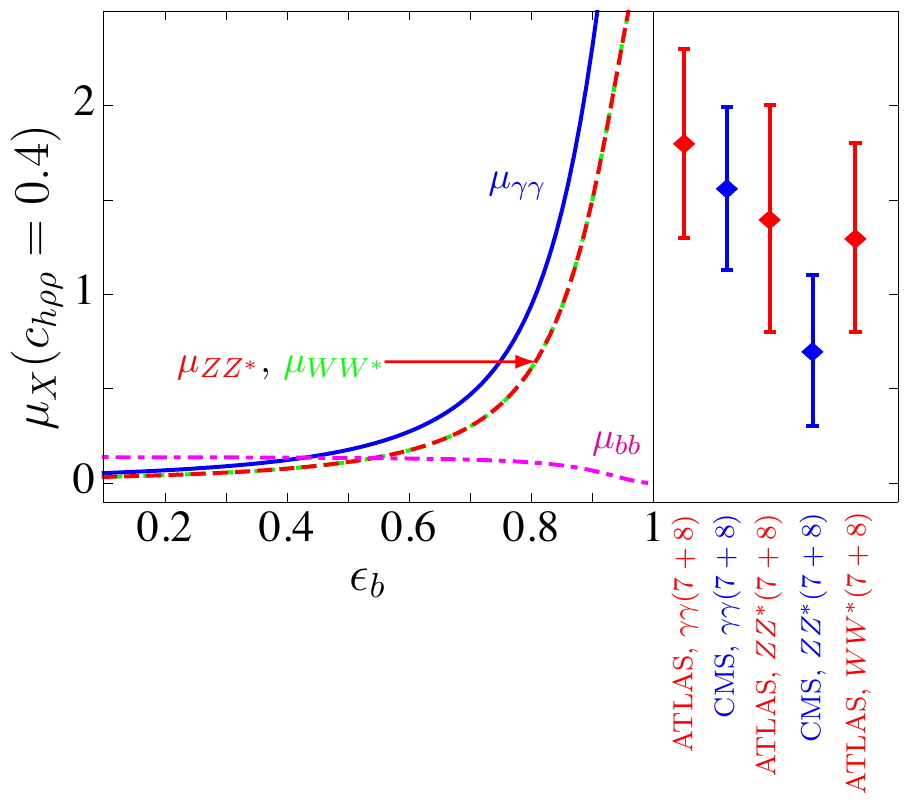} 
\vspace*{2ex}
\end{minipage}
}
\end{tabular}
\caption[]{
The signal strength $\mu_X (X = \gamma\gamma,WW^*,ZZ^*,bb)$ as a unction of $\epsilon_b$ in the present model. The blue solid, green dotted, red dashed, magenta dot-dashed curves correspond to $\mu_{\gamma \gamma}, \mu_{WW^*},\mu_{ZZ^*}, \mu_{bb}$, respectively. In both panels, the LHC combined results for $\gamma \gamma\,,\,ZZ^*\,,\,WW^*$ in Eqs.(\ref{signalstrength-ATLAS},\ref{signalstrength-CMS}) are shown together. (a) shows the signal strength for a case with $C_{h\rho\rho} = 0$ and (b) shows the signal strength for a case with $C_{h\rho\rho} = 0.4$ with $M_\rho = 1 \TeV$ 
\label{higgs-LHC}}
\end{center}
\end{figure}

In Fig.\ref{higgs-LHC} (a), we show the signal strength $\mu_X$ as a function of $\epsilon_b$. In Fig.\ref{higgs-LHC} (a), the blue solid, green dotted, red dashed and magenta dot-dashed curves correspond to $\mu_{\gamma \gamma}, \mu_{WW^*},\mu_{ZZ^*}, \mu_{bb}$, respectively. Moreover, for comparison, we present  the values of $\mu_{\gamma \gamma , ZZ^* , WW^*}$ corresponding to the results reported by the ATLAS  and CMS experiments and given in Eqs.(\ref{signalstrength-ATLAS}) and (\ref{signalstrength-CMS}). From Fig.\ref{higgs-LHC} (a), we find $\epsilon_b = 0.7 - 0.93$ is favored by the experimental data on $\mu_{WW^*,ZZ^*}$
When $\epsilon_b$ becomes large, $s^b_R$ becomes small. Consequentially, $C_{hbb}$ in Eq.(\ref{chbb}) becomes small and ${\rm Br}(h \to b\bar{b})$ becomes small for large $\epsilon_b$. This fact causes the enhancement of ${\rm Br}(h \to WW^*/ZZ^*)$ for large $\epsilon_b$. For  $\epsilon_b = 0.7 - 0.93$, we obtain $\mu_{bb} = 0.12 - 0.04$ which is smaller than the SM higgs boson case but still consistent with the LHC results. 
However, the $\mu_{\gamma \gamma}$ remains smaller than the LHC results even if we take into account the effect of $\epsilon_b$. 

To conclude this section, we discuss a possibility of enhancing $\mu_{\gamma\gamma}[\text{TSSTC}] \simeq 2$ while retaining the features of the other channels. For this purpose, there are three possibilities :
\begin{enumerate}
\item Adding new vector mesons which couple to higgs boson,
\item Adding new fermions which couple to higgs boson,
\item Adding new scalar particles which couple to higgs boson.
\end{enumerate}
Among these possibilities, the first one occurs naturally in the present model, since the topcolor dynamics generates composite vector mesons. Let us denote 
such color-singlet vector meson isotriplet by $\rho^\pm_\mu$, and assume its mass to satisfy 
$M_\rho \gg 2 m_h$. We add 
\beq
{\cal L}_{h\rho\rho}
= C_{h\rho\rho} (g M_W) \cdot h \rho^{+ \mu} \rho^-_\mu 
\eeq 
to Eq.(\ref{Lag-hdecay}). In this case $\Gamma(h \to \gamma \gamma)$ changes from 
Eq.(\ref{1loop-gammagamma}) to
\beq
\Gamma(h \to \gamma \gamma)
\!\!&=&\!\!
\frac{\alpha^2 g^2}{1024 \pi^3}\frac{m^3_h}{M^2_W}
\left| 
C_{hWW} A_1\left(\frac{4M^2_W}{m^2_h} \right)
+
C_{h\rho\rho} A_1\left(\frac{4M^2_\rho}{m^2_h} \right)
+ \cdots
\right|^2
\,,\label{1loop-gammagamma-newV}
\eeq
where $\cdots$ contain the $A_{1/2,0}$-terms in Eq.(\ref{1loop-gammagamma}). If $M_\rho \gg 2 m_h \simeq 2 \times 126 \GeV$, $A_1$ can be take to be equal to  $-7$. Then $\Gamma(h \to \gamma \gamma)$ can be enhanced for suitable values of $C_{h\rho\rho}$. Furthermore, this new vector meson does not give any contribution to the other decay channels at the leading order since $\rho^\pm_\mu$ is color-singlet and $M_\rho \gg 2 m_h$. %
In Fig.\ref{higgs-LHC}(b), we show the signal strength $\mu_X$ for $C_{h\rho\rho}$, with 
$M_\rho = 1 \TeV$, as a function of $\epsilon_b$. In Fig.\ref{higgs-LHC} (b), the blue solid, green dotted, red dashed and magenta dot-dashed curves correspond to $\mu_{\gamma \gamma}, \mu_{WW^*},\mu_{ZZ^*}, \mu_{bb}$, respectively. For comparison, we again also present  the values of 
$\mu_{\gamma \gamma , ZZ^* , WW^*}$ from the ATLAS 
and  CMS experiments. 
The dependence of the results on $M_\rho$ is small for $M_\rho \gg 2m_h$ due to the loop function $A_1(x)$. 
Summarizing, we find that this modification, i.e. adding ${\cal L}_{h\rho\rho}$, gives a large contribution to $\Gamma(h \to \gamma \gamma)$ but $\text{Br}(h \to WW^*/ZZ^*/b\bar{b})$ are not affected by this addition. 
Therefore, from Fig.\ref{higgs-LHC}, we find that the present model with 
\beq
\epsilon_b = 0.7 - 0.93 \quad \text{and} \quad C_{h \rho \rho} \simeq 0.4
\,,
\eeq
is consistent with the experimental constraints and the LHC results of higgs boson search. Future data from the LHC will allow to constrain the model further. Especially interesting will be the fate of the deficit observed in the $\tau\tau$-channel, and which is by definition explained within our model. If a signal
in the $\tau\tau$-channel is ultimately observed, the model must be revised to accommodate such a result.

%
\section{Summary}
%
In this paper we have explored a model where both electroweak symmetry breaking 
and the origin of the heavy quark masses are due to new strong dynamics. In the model
we considered, the third generation quark masses arise from the topcolor interactions via the
top-seesaw mechanism. These augment a technicolor sector which is mainly responsible 
for the generation of the masses of the weak interaction gauge bosons.

The resulting low energy effective theory is a particular three Higgs doublet model. Several 
novel properties were identified. We considered the CP even scalar state associated with the 
technicolor sector to be heavy. This assumption is reasonable for the minimalistic 
technicolor sector we considered, but may be alleviated for other possibilities. In particular, if 
the technicolor sector is quasiconformal, the scalar state is expected to be light and contribute 
to the mass eigenstates in the scalar sector.

In the phenomenology analysis of the model we have provided a template on how to 
confront this type of models with the existing data from the precision electroweak measurements
to the recently announced LHC discovery results. In particular we find that a natural way to 
accommodate the possible observed enhancement in the $h_0\rightarrow \gamma\gamma$ 
channel, is via the composite vector state which inevitably exist in this type of models.

Our analysis of the model parameter space is to be taken as only illustrative. We have provided the
necessary formulas and concepts, shown that viable portion in the parameter space exists and 
laid out the way for the more detailed scan of the parameter space. The future results from the LHC
on the fate of the excess in the $\gamma\gamma$ channel as well as on the deficit in the $\tau\tau$
final states will certainly provide stringent constraints on this type of models.

\appendix
\section{Results for the analysis of the oblique corrections and $\delta g_L^b$}
\label{ewptappendix}

From Eq. (\ref{hybrid-full-EFT}) in the mass basis of PNGBs and higgses, the following
Feynman rules are obtained:

\begin{table}[h]
{
\tabcolsep=1ex
\renewcommand\arraystretch{1}
\begin{tabular}{|c|c|}
\hline 
$ SVV$-vertex & Feynman rule
\\ \hline \hline 
$H^0 Z_\mu Z_\nu$ & $i (g/c_W)M_Z \cos \phi \cos (\beta - \alpha)g^{\mu\nu}$
\\\hline 
$h^0 Z_\mu Z_\nu$ & $i(g/c_W)M_Z \cos \phi \sin (\beta - \alpha)g^{\mu\nu}$
\\\hline
$G^\pm A_\mu W^\mp_\nu$ & $igs_W M_W g^{\mu\nu}$
\\\hline 
$G^\pm Z_\mu W^\mp_\nu$ & $-igs^2_WM_Z g^{\mu\nu}$
\\\hline
$H^0 W^+_\mu W^-_\nu$ & $ig M_W \cos \phi \cos(\beta - \alpha) g^{\mu\nu}$
\\\hline
$h^0 W^+_\mu W^-_\nu$ & $ig M_W \cos \phi \sin(\beta - \alpha) g^{\mu\nu}$
\\\hline 
\end{tabular}
}
\caption{
Feynman rules for $SVV$-type  vertices for Fig.\ref{higgs-VP}(a).
\label{FR-SVV}
}
\end{table}
\begin{table}[h]
{
\tabcolsep=1ex
\renewcommand\arraystretch{1,3}
\begin{tabular}{|c|c|}
\hline 
$SS V$-vertex & Feynman rule
\\ \hline \hline 
$S^+ S^- A^\mu \,,(S=G,H_{1,2})$ & $i e (p_+ - p_-)^\mu$
\\\hline
$S^+ S^- Z^\mu \,,(\phi=G,H_{1,2})$ & $i[g(c^2_W-s^2_W)/(2c_W)] (p_+ - p_-)^\mu$ 
\\\hline
$G^0 H^0 Z^\mu $ & $-g/(2c_W) \cos \phi \cos(\beta-\alpha) (p_G - p_H)^\mu$
\\\hline
$G^0 h^0 Z^\mu $ & $-g/(2c_W) \cos \phi \sin(\beta-\alpha) (p_G - p_h)^\mu$
\\\hline
$A^0_1 H^0 Z^\mu $ & $-g/(2c_W) [-\sin \zeta_0 \sin(\beta- \alpha) - \sin \phi \cos \zeta_0 \cos (\beta-\alpha)] (p_A - p_H)^\mu$
\\\hline
$A^0_1 h^0 Z^\mu $ & $-g/(2c_W) [\sin \zeta_0 \cos(\beta- \alpha) - \sin \phi \cos \zeta_0 \sin (\beta-\alpha)] (p_A - p_h)^\mu$
\\\hline
$A^0_2 h^0 Z^\mu $ & $-g/(2c_W) [\cos \zeta_0 \cos(\beta- \alpha) + \sin \phi \sin \zeta_0 \sin (\beta-\alpha)] (p_A - p_h)^\mu$
\\\hline
$A^0_2 H^0 Z^\mu $ & $-g/(2c_W) [-\cos \zeta_0 \sin(\beta- \alpha) + \sin \phi \sin \zeta_0 \cos (\beta-\alpha)] (p_A - p_H)^\mu$
\\\hline
$G^\pm H^0 W^{\mp \mu} $ & $ \pm i (g/2) \cos \phi \cos(\beta-\alpha) (p_G - p_H)^\mu$
\\\hline
$G^\pm h^0 W^{\mp \mu} $ & $ \pm i (g/2) \cos \phi \sin(\beta-\alpha) (p_G - p_h)^\mu$
\\\hline
$H^\pm_1 H^0 W^{\mp \mu} $ & $\pm i (g/2) [-\sin \zeta_\pm \sin(\beta- \alpha) - \sin \phi \cos \zeta_\pm \cos (\beta-\alpha)] (p_{H^\pm} - p_H)^\mu$
\\\hline
$H^\pm_1 h^0 W^{\mp \mu}$ & $\pm i (g/2) [\sin \zeta_\pm \cos(\beta- \alpha) - \sin \phi \cos \zeta_\pm \sin (\beta-\alpha)] (p_{H^\pm} - p_h)^\mu$
\\\hline
$H^\pm_2 H^0 W^{\mp \mu} $ & $\pm i (g/2) [-\cos \zeta_\pm \sin(\beta- \alpha) + \sin \phi \sin \zeta_\pm \cos (\beta-\alpha)] (p_{H^\pm} - p_H)^\mu$
\\\hline
$H^\pm_2 h^0 W^{\mp \mu}$ & $\pm i (g/2) [\cos \zeta_\pm \cos(\beta- \alpha) + \sin \phi \sin \zeta_\pm \sin (\beta-\alpha)] (p_{H^\pm} - p_h)^\mu$
\\\hline
$G^\pm G^0 W^{\mp\mu}$ & $-(g/2) (p_{G^\pm} - p_{G^0})$
\\\hline
$H^\pm_1 A^0_1 W^{\mp\mu}$ & $-(g/2) \cos (\zeta_\pm - \zeta_0)(p_{H^\pm} - p_A)$
\\\hline
$H^\pm_1 A^0_2 W^{\mp\mu}$ & $-(g/2) \sin (\zeta_\pm - \zeta_0)(p_{H^\pm} - p_A)$
\\\hline
$H^\pm_2 A^0_1 W^{\mp\mu}$ & $(g/2) \sin (\zeta_\pm - \zeta_0)(p_{H^\pm} - p_A)$
\\\hline
$H^\pm_2 A^0_2 W^{\mp\mu}$ & $-(g/2) \cos (\zeta_\pm - \zeta_0)(p_{H^\pm} - p_A)$
\\\hline
\end{tabular}
}
\caption{
Feynman rules for $SSV$-type vertices for Fig.\ref{higgs-VP}(b). The four-momentum $p_i$ points into the vertex.
\label{FR-SSV}
}
\end{table}
\begin{table}[b]
{
\tabcolsep=1ex
\renewcommand\arraystretch{1.3}
\begin{tabular}{|c|c|}
\hline 
$SSVV$-vertex & Feynman rule
\\ \hline \hline 
$S^0 S^0 Z^\mu Z^\nu \,,(S=h,H,G,A_{1,2})$ & $i (g^2/2c^2_W)g^{\mu\nu}$
\\\hline
$S^+ S^- Z^\mu Z^\nu \,,(S=G,H_{1,2})$ & $i [g^2(c^2_W - s^2_W)/(2c^2_W)] g^{\mu\nu}$
\\\hline
$S^+ S^- A^\mu A^\nu \,,(S=G,H_{1,2})$ & $2 i e^2 g^{\mu\nu}$
\\\hline
$S^+ S^- A^\mu Z^\nu \,,(S=G,H_{1,2})$ & $i [g^2s_W(c^2_W - s^2_W)/c_W] g^{\mu\nu}$
\\\hline
$S^0 S^0 W^{+\mu}W^{-\nu} \,,(S=h,H,G,A_{1,2})$ & $i (g^2/2)g^{\mu\nu}$
\\\hline
$S^+ S^- W^{+\mu}W^{-\nu} \,,(S=G,H_{1,2})$ & $i (g^2/2)g^{\mu\nu}$
\\\hline
\end{tabular}
}
\caption{
Feynman rules for $SSVV$-type vertices for Fig.\ref{higgs-VP}(c) which contribute to the Peskin-Takeuchi $S,T$-parameters. 
\label{FR-SSVV}
}
\end{table}%

Computing the relevant diagrams corresponding to  Fig.\ref{higgs-VP},
the Peskin-Takeuchi $S$ parameter for the higgs sector in the present model is given by
\beq
S
=
\frac{1}{ \pi M^2_Z}
\left[
\begin{aligned}
&
\cos^2\phi \cos^2(\beta-\alpha)
\left\{
\begin{aligned}
&
{\cal B}_{00} (M^2_Z,M^2_Z,m^2_H)
-{\cal B}_{00} (M^2_Z,M^2_Z,m^2_h)
\\
&
-M^2_Z {\cal B}_0(M^2_Z,m^2_H,M^2_Z)
+M^2_Z {\cal B}_0(M^2_Z,m^2_h,M^2_Z)
\end{aligned}
\right\}
\\[1ex]
&
- {\cal B}_{00} (M^2_Z,m^2_{H^\pm_1},m^2_{H^\pm_1})
- {\cal B}_{00} (M^2_Z,m^2_{H^\pm_2},m^2_{H^\pm_2})
-
\sin^2\phi {\cal B}_{00}(M^2_Z,M^2_Z,m^2_h)
\\[0.3ex]
&
+ (\sin \zeta_0 \sin(\beta-\alpha) + \sin \phi \cos \zeta_0 \cos(\beta-\alpha))^2
{\cal B}_{00}(M^2_Z,m^2_{A_1},m^2_H)
\\[0.3ex]
&
+ (\sin \zeta_0 \cos(\beta-\alpha) - \sin \phi \cos \zeta_0 \sin(\beta-\alpha))^2
{\cal B}_{00}(M^2_Z,m^2_{A_1},m^2_h)
\\[0.3ex]
&
+ (\cos \zeta_0 \sin(\beta-\alpha)- \sin \phi \sin \zeta_0 \cos(\beta-\alpha))^2
{\cal B}_{00}(M^2_Z,m^2_{A_2},m^2_H)
\\[0.3ex]
&
+ (\cos \zeta_0 \cos(\beta-\alpha) + \sin \phi \sin \zeta_0 \sin(\beta-\alpha))^2
{\cal B}_{00}(M^2_Z,m^2_{A_2},m^2_h)
\\[0.3ex]
\end{aligned}
 \right]
\,.\label{S-full}
\eeq
Similarly, the Peskin-Takeuchi $T$ parameter for the higgs sector in the model is given as
\beq
T
=
\frac{1}{4\pi M^2_W s^2_W}
\left[ 
\begin{aligned}
&
\cos^2 (\zeta_\pm - \zeta_0)
\left\{ 
B_{00}(0,m^2_{H^\pm_1},m^2_{A_1}) + B_{00}(0,m^2_{H^\pm_2},m^2_{A_2}) 
\right\}
\\[1ex]
&
+
\sin^2 (\zeta_\pm - \zeta_0)
\left\{ 
B_{00}(0,m^2_{H^\pm_1},m^2_{A_2}) + B_{00}(0,m^2_{H^\pm_2},m^2_{A_1}) 
\right\}
\\[1ex]
&
+ \cos^2\phi \cos^2(\beta-\alpha)
\left\{ 
\begin{aligned}
& B_{00}(0,M^2_W,m^2_H) - B_{00}(0,M^2_Z,m^2_H) \\[0.3ex]
& -B_{00}(0,M^2_W,m^2_h) + B_{00}(0,M^2_Z,m^2_h)
\end{aligned}
\right\}
\\[1ex]
&
-\sin^2\phi 
\left\{ B_{00}(0,M^2_W,m^2_h) - B_{00}(0,M^2_Z,m^2_h)\right\}
\\[1ex]
&
-M^2_W \cos^2 \phi \cos^2(\beta-\alpha) 
\left\{ B_0(0,m^2_H,M^2_W) - B_0(0,m^2_h,M^2_W)\right\}
\\[1ex]
&
+M^2_Z \cos^2 \phi \cos^2(\beta-\alpha) 
\left\{ B_0(0,m^2_H,M^2_Z) - B_0(0,m^2_h,M^2_Z)\right\}
\\[1ex]
&
- \frac{1}{2} \left\{  A_0 (m^2_{H^\pm_1}) + A_0 (m^2_{H^\pm_2}) \right\}
\\[1ex]
&
+ (\sin \zeta_\pm \sin(\beta-\alpha)+ \sin \phi \cos \zeta_\pm \cos(\beta-\alpha))^2
B_{00}(0,m^2_{H^\pm_1},m^2_H)
\\[0.3ex]
&
+ (\sin \zeta_\pm \cos(\beta-\alpha)- \sin \phi \cos \zeta_\pm \sin(\beta-\alpha))^2
B_{00}(0,m^2_{H^\pm_1},m^2_h)
\\[0.3ex]
&
+ (\cos \zeta_\pm \sin(\beta-\alpha)- \sin \phi \sin \zeta_\pm \cos(\beta-\alpha))^2
B_{00}(0,m^2_{H^\pm_2},m^2_H)
\\[0.3ex]
&
+ (\cos \zeta_\pm \cos(\beta-\alpha)+ \sin \phi \sin \zeta_\pm \sin(\beta-\alpha))^2
B_{00}(0,m^2_{H^\pm_2},m^2_h)
\\[0.3ex]
&
-(\sin \zeta_0 \sin(\beta-\alpha)+ \sin \phi \cos \zeta_0 \cos(\beta-\alpha))^2
B_{00}(0,m^2_{A_1},m^2_H)
\\[0.3ex]
&
-(\sin \zeta_0 \cos(\beta-\alpha)- \sin \phi \cos \zeta_0 \sin(\beta-\alpha))^2
B_{00}(0,m^2_{A_1},m^2_h)
\\[0.3ex]
&
-(\cos \zeta_0 \sin(\beta-\alpha)- \sin \phi \sin \zeta_0 \cos(\beta-\alpha))^2
B_{00}(0,m^2_{A_2},m^2_H)
\\[0.3ex]
&
-(\cos \zeta_0 \cos(\beta-\alpha)+ \sin \phi \sin \zeta_0 \sin(\beta-\alpha))^2
B_{00}(0,m^2_{A_2},m^2_h)
\end{aligned}
\right]
\,,\label{T-full}
\eeq
where we compute in the 'tHoot-Feynman gauge with dimensional regularization and $A_0,B_{00},B_0,{\cal B}_{00},{\cal B}_0$ are given by
\beq
A_0(m^2)
\!\!&=&\!\!
 m^2 \left[ \frac{1}{\bar{\epsilon}} + 1 - \ln m^2\right]
 \,,
\\[1ex]
B_{00}(q^2,m^2_1,m^2_2)
\!\!&=&\!\!
\left(\frac{m^2_1 + m^2_2}{4} -\frac{1}{12}q^2\right) 
\left(\frac{1}{\bar{\epsilon}}+1\right)
-\frac{1}{2}\int^1_0\!\!dx \Delta\ln \Delta
\,,\\[1ex]
B_0(q^2,m^2_1,m^2_2)
\!\!&=&\!\!
\frac{1}{\bar{\epsilon}}
- \int^1_0\!\!dx \ln \Delta
\,,\\[1ex]
\Delta 
\!\!\!&\equiv&\!\!\! 
-x(1-x)q^2 + (1-x)m^2_1 + x m^2_2\,,
\\[1ex]
\frac{1}{\bar{\epsilon}}
\!\!\!&\equiv&\!\!\!
\frac{2}{4-d} - \gamma_E + \ln (4\pi)
\,,\\[1ex]
{\cal B}_{00}(q^2,m^2_1,m^2_2)
\!\!&\equiv&\!\! 
B_{00}(q^2,m^2_1,m^2_2) - B_{00}(0,m^2_1,m^2_2)
\,,\\[1ex]
{\cal B}_0(q^2,m^2_1,m^2_2)
\!\!&\equiv&\!\! 
B_0(q^2,m^2_1,m^2_2) - B_0(0,m^2_1,m^2_2)
\,.
\eeq

We are computing using dimensional regularization, and we remark that these results are almost the 
same as the results for the three higgs doublet model. However, the difference is that one higgs 
doublet among  three higgs doublets does not have the CP-even higgs boson; 
see Eq. (\ref{TC-higgs}). This implies 
that Eqs. (\ref{S-full}) and (\ref{T-full}) have a divergent part proportional to 
$1/\bar{\epsilon}$ since we treat the TC sector by using the non-linear sigma model. 

Note that if we take a limit, $\tan \phi =0$, $c_1 = c_2 =0$, i.e. $\cos \zeta_p = 0$, and $m^2_{S_2} \gg M^2_Z,m^2_{S_1}$, Eqs.(\ref{S-full},\ref{T-full}) becomes 
\beq
 S
=
\frac{1}{ \pi M^2_Z}
\left[
\begin{aligned}
&
\cos^2(\beta-\alpha)
\left\{
\begin{aligned}
&
{\cal B}_{00} (M^2_Z,M^2_Z,m^2_H)
-{\cal B}_{00} (M^2_Z,M^2_Z,m^2_h)
\\
&
-M^2_Z {\cal B}_0(M^2_Z,m^2_H,M^2_Z)
+M^2_Z {\cal B}_0(M^2_Z,m^2_h,M^2_Z)
\\
&
+{\cal B}_{00}(M^2_Z,m^2_{A_H},m^2_h)
\end{aligned}
\right\}
\\[1ex]
&
- {\cal B}_{00} (M^2_Z,m^2_{H^\pm_H},m^2_{H^\pm_H})
+ \sin^2(\beta-\alpha)
{\cal B}_{00}(M^2_Z,m^2_{A_H},m^2_H)
\end{aligned}
 \right]
\,,
\eeq
and
\beq
 T
=
\frac{1}{4\pi M^2_W s^2_W}
\left[ 
\begin{aligned}
&
B_{00}(0,m^2_{H^\pm_H},m^2_{A_H})
- \frac{1}{2}  A_0 (m^2_{H^\pm_L}) 
\\[1ex]
&
+ \cos^2(\beta-\alpha)
\left\{ 
\begin{aligned}
& B_{00}(0,M^2_W,m^2_H) - B_{00}(0,M^2_Z,m^2_H) \\[0.3ex]
& -B_{00}(0,M^2_W,m^2_h) + B_{00}(0,M^2_Z,m^2_h) \\[0.3ex]
& +B_{00}(0,m^2_{H^\pm_H},m^2_h) - B_{00}(0,m^2_{A_H},m^2_h)
\end{aligned}
\right\}
\\[1ex]
&
+ \sin^2(\beta-\alpha)
\left\{
B_{00}(0,m^2_{H^\pm_H},m^2_H)
- B_{00}(0,m^2_{A_H},m^2_H)
\right\}
\\[1ex]
&
-M^2_W \cos^2(\beta-\alpha) 
\left\{ B_0(0,m^2_H,M^2_W) - B_0(0,m^2_h,M^2_W)\right\}
\\[1ex]
&
+M^2_Z \cos^2(\beta-\alpha) 
\left\{ B_0(0,m^2_H,M^2_Z) - B_0(0,m^2_h,M^2_Z)\right\}
\end{aligned}
\right]
\,,
\eeq
which are finite and reproduce the 2HDM results \cite{He:2001fz,He:2001tp} as they should.

\begin{table}[htbp]
\tabcolsep=1ex
\renewcommand\arraystretch{2}
\begin{tabular}{|c|c||c|c|}
\hline 
 operator &Coupling strength & operator & Coupling strength
\\ \hline
$Z_\mu \bar{t}_L \gamma^\mu t_L $ & $\dfrac{g}{c_W} \left[ g^t_L - \dfrac{1}{2} (s^t_L)^2 \right]$ 
&
$Z_\mu \bar{b}_L \gamma^\mu b_L $ & $\dfrac{g}{c_W} \left[ g^b_L + \dfrac{1}{2} (s^b_L)^2\right]$ 
\\ \hline
$Z_\mu \left(\bar{t}_L \gamma^\mu T_L  + \bar{T}_L \gamma^\mu t_L\right)$ & $\dfrac{g}{c_W} \left[ \dfrac{1}{2} c^t_L s^t_L \right]$ 
&
$Z_\mu \left(\bar{b}_L \gamma^\mu B_L  + \bar{B}_L \gamma^\mu b_L\right)$ & $\dfrac{g}{c_W} \left[ -\dfrac{1}{2} c^b_L s^b_L\right]$ 
\\ \hline
$Z_\mu \bar{T}_L \gamma^\mu T_L $ & $\dfrac{g}{c_W} \left[ g^t_L - \dfrac{1}{2} (c^t_L)^2\right]$  
&
$Z_\mu \bar{B}_L \gamma^\mu B_L $ & $\dfrac{g}{c_W} \left[ g^b_L + \dfrac{1}{2} (c^b_L)^2\right]$  
\\ \hline 
$Z_\mu \left(\bar{t}_R \gamma^\mu t_R  + \bar{T}_R \gamma^\mu T_R\right)$ & $\dfrac{g}{c_W} \left[ g^t_R \right]$  
&
$Z_\mu \left(\bar{b}_R \gamma^\mu b_R  + \bar{B}_R \gamma^\mu B_R\right)$ & $\dfrac{g}{c_W} \left[ g^b_R \right]$  
\\ \hline \hline
$W^+_\mu \bar{t}_L \gamma^\mu b_L  + \text{h.c.}$ & $\dfrac{g}{\sqrt{2}} \left[ c^t_L c^b_L \right]$  
&
$W^+_\mu \bar{t}_L \gamma^\mu B_L  + \text{h.c.}$ & $\dfrac{g}{\sqrt{2}} \left[ c^t_L s^b_L \right]$  
\\ \hline
$W^+_\mu \bar{T}_L \gamma^\mu b_L  + \text{h.c.}$ & $\dfrac{g}{\sqrt{2}} \left[ s^t_L c^b_L \right]$  
&
$W^+_\mu \bar{T}_L \gamma^\mu B_L  + \text{h.c.}$ & $\dfrac{g}{\sqrt{2}} \left[ s^t_L s^b_L \right]$  
\\ \hline
\end{tabular}
\caption{
$V\bar{f}f$-couplings.
\label{Vff-couplings}
}
\end{table}

\begin{table}[tb]
\begin{tabular}{cc}
{
\tabcolsep=1ex
\renewcommand\arraystretch{2}
\begin{tabular}{|c|c|}
\hline 
 operator &Coupling strength
 \\ \hline
%
%
%
%
%
$G^+ \bar{t}_R b_L + \text{h.c.}$ & $Y^G_{tb} = y_2 s^t_R c^b_L \cos \phi \sin \beta + y^t_{\rm TC} \sin \phi$ 
\\ \hline
$H^+_L \bar{t}_R b_L + \text{h.c.}$ & $Y^L_{tb} =  y_2 s^t_R c^b_L (\cos \beta \cos \zeta_\pm + \sin \phi \sin \beta \sin \zeta_\pm) - y^t_{\rm TC} \cos \phi \sin \zeta_\pm$ 
\\ \hline
$H^+_H \bar{t}_R b_L + \text{h.c.}$ & $Y^H_{tb} =  y_2 s^t_R c^b_L(\cos \beta \sin \zeta_\pm - \sin \phi \sin \beta \cos \zeta_\pm) + y^t_{\rm TC} \cos \phi \cos \zeta_\pm$ 
\\ \hline \hline
%
%
$G^+ \bar{T}_R b_L + \text{h.c.}$ & $ Y^G_{Tb} = y_2 c^t_R c^b_L \cos \phi \sin \beta $ 
\\ \hline
$H^+_L \bar{T}_R b_L + \text{h.c.}$ & $ Y^L_{Tb} =  y_2 c^t_R c^b_L (\cos \beta \cos \zeta_\pm + \sin \phi \sin \beta \sin \zeta_\pm)$ 
\\ \hline
$H^+_H \bar{T}_R b_L + \text{h.c.}$ & $ Y^H_{Tb} =  y_2 c^t_R c^b_L(\cos \beta \sin \zeta_\pm - \sin \phi \sin \beta \cos \zeta_\pm) $ 
%
\\ \hline 
\end{tabular}
}
\end{tabular}
\caption{
Couplings between charged scalars and left-handed bottom quark.
\label{charged-Gff-couplings}
}
\end{table}

The one-loop corrections to $\delta g_L^b$ are obtained as

\beq
[\delta g^b_L]^{\text{1loop}}_{\text{gauge}} 
\!\!&=&\!\!
\frac{1}{2}
\frac{g^2 }{16 \pi^2}  [c^b_L]^2 
\Big(
[c^t_L]^2 C_{01}(m^2_t,M^2_W) 
+ [s^t_L]^2 C_{01}(m^2_T,M^2_W) 
\Big)
+
\frac{g^2 c^2_W}{16 \pi^2}  [c^b_L]^2 
\Big(
 C_{001}(m^2_t,M^2_W) 
+ C_{001}(m^2_T,M^2_W) 
\Big)
\,\nonumber\\[1ex]
&&
-
\frac{g^2}{64 \pi^2}  [c^b_L]^2  [s^b_L]^2
\left(
[c^t_L]^2 B(m^2_t,M^2_W)
+
[s^t_L]^2 B(m^2_T,M^2_W)
\right)
+
\frac{g^2}{64 \pi^2}[c^b_L]^2 [c^t_L]^2  [s^t_L]^2 C_{004}(m^2_t,m^2_T,M^2_W)
\,\nonumber\\[1ex]
&&
-
\frac{g s^2_W }{8 \pi^2 \sqrt{2}}  [c^b_L] \cdot 
\left( 
m_t M_W  [c^t_L][Y^G_{tb}] C_{02} (m^2_t,M^2_W)
+
m_T M_W  [s^t_L][Y^G_{Tb}] C_{02} (m^2_T,M^2_W)
\right)
\,,
\label{Zbb-1loop-gauge-result}
\eeq
and 
\beq
[\delta g^b_L]^{\text{1loop}}_{\text{NGB}} 
\!\!&=&\!\!
- \frac{1}{2}\frac{1}{16 \pi^2}  
\sum_{\{f\}}\sum_{\{h\}} [Y^h_{fb}]^2 C_{01}(m^2_f,M^2_h)
-
\frac{1}{64 \pi^2}  [s^b_L]^2
\sum_{\{f\}}\sum_{\{h\}} [Y^h_{fb}]^2 B_{01}(m^2_f,M^2_h)
\,\nonumber\\[1ex]
&&
+\frac{1}{2}\frac{1}{16 \pi^2} [s^t_L]^2  
\sum_{\{h\}} [Y^h_{tb}]^2 C_{01}(m^2_t,M^2_h)
+\frac{1}{2}\frac{1}{16 \pi^2} [c^t_L]^2  
\sum_{\{h\}} [Y^h_{Tb}]^2 C_{01}(m^2_T,M^2_h)
\,,\nonumber\\[1ex]
&&
+\frac{1}{2}\frac{1}{16 \pi^2}  [s^t_L c^t_L]
\sum_{\{h\}} [Y^h_{tb}][Y^h_{Tb}] C_{03}(m^2_t,m^2_T,M^2_h)
\,,
\label{Zbb-1loop-NGB-only-result}
\eeq
where $\{ f \} = t,T$ and $\{ h \} = G^\pm,H^\pm_L,H^\pm_H$ and 
\beq
B(m^2,M^2)
\!\!&=&\!\!
-\frac{m^2}{M^2 - m^2} - \frac{m^4}{(M^2-m^2)^2}\ln \frac{m^2}{M^2}
\,,\\[1ex]
C_{001}(m^2,M^2)
\!\!&=&\!\!
\frac{m^2}{M^2 - m^2} + \frac{m^4}{(M^2-m^2)^2}\ln \frac{m^2}{M^2}
\,,\\[1ex]
C_{002}(m^2,M^2)
\!\!&=&\!\!
-\frac{m^2}{M^2 - m^2} + \frac{m^2(m^2 - 2 M^2)}{(M^2-m^2)^2}\ln \frac{m^2}{M^2}
\,,\\[1ex]
C_{003}(m^2_1,m^2_2,M^2)
\!\!&=&\!\!
\frac{1}{m^2_1 - m^2_2}
\left[
\frac{m^2_1(2M^2 - m^2_1)}{m^2_1 - M^2} \ln \frac{m^2_1}{M^2}
- \frac{m^2_2 (2M^2 -m^2_2)}{m^2_2 - M^2} \ln \frac{m^2_2}{M^2}
\right]
\,,\\[1ex]
C_{004}(m^2_1,m^2_2,M^2)
\!\!&=&\!\!
C_{002}(m^2_1,M^2) + C_{002}(m^2_2) + 2 C_{003}(m^2_1,m^2_2,M^2)
\,,\\[1ex]
C_{01}(m^2,M^2)
\!\!&=&\!\!
\frac{m^2}{M^2 - m^2} + \frac{m^2 M^2}{(M^2-m^2)^2}\ln \frac{m^2}{M^2}
\,,\\[1ex]
C_{02}(m^2,M^2)
\!\!&=&\!\!
\frac{1}{M^2 - m^2} - \frac{m^2}{(M^2-m^2)^2}\ln \frac{m^2}{M^2}
\,,\\[1ex]
C_{03}(m^2_1,m^2_2,M^2)
\!\!&=&\!\!
\frac{-m_1 m_2}{m^2_1 - m^2_2}
\left[
\frac{m^2_1}{m^2_1 - M^2} \ln \frac{m^2_1}{M^2}
- \frac{m^2_2}{m^2_2 - M^2} \ln \frac{m^2_2}{M^2}
\right]
\,.
\eeq

\section{Decay widths of the lightest CP even scalar in the model}
\label{phenoappendix}

The two body decay width $\Gamma(h \to WW/ZZ/\bar{f}f)$ are given by \cite{Gunion:1989we}
\beq
\Gamma(h \to WW) 
\!\!&=&\!\!
 |C_{hWW}|^2
\frac{g^2}{64 \pi} \frac{m^3_h}{M^2_W} \sqrt{1- \frac{4 M^2_W}{m^2_h}} \left[ 1 - \frac{4 M^2_W}{m^2_h} + \frac{16 M^4_W}{m^4_h}\right]\,,
\quad (\text{for $4 M^2_W \leq m^2_h$})
\,,\label{2decay-WW}\\[1ex]
\Gamma(h \to ZZ) 
\!\!&=&\!\!
 |C_{hWW}|^2
\frac{g^2}{128 \pi} \frac{m^3_h}{M^2_W}  \sqrt{1- \frac{4 M^2_Z}{m^2_h}} \left[ 1 - \frac{4 M^2_Z}{m^2_h} + \frac{16 M^4_Z}{m^4_h}\right]\,,
\quad (\text{for $4 M^2_Z \leq m^2_h$})
\,,\label{2decay-ZZ}\\[1ex]
\Gamma(h \to ff) 
\!\!&=&\!\!
 |C_{hff}|^2
\frac{3 g^2}{32 \pi} \frac{m^2_f}{M^2_W} m_h \left[1- \frac{4 m^2_f}{m^2_h} \right]^{3/2} \,,
\quad (\text{for $4 m^2_f \leq m^2_h$})
\,, \label{2decay-ff}
\eeq 
the three body decay width $\Gamma(\phi \to WW^*/ZZ^*)$ are given by \cite{Rizzo:1980gz}
\beq
\Gamma(h \to W W^*)
\!\!&=&\!\!
|C_{hWW}|^2 \left[ 3+ C^2_{Wtb} \right]
\frac{g^4 m_h}{512\pi^3} F\left( \frac{M_W}{m_h}\right)
\,,
\quad (\text{for $ M_W \leq m_h \leq 2 M_W$}) 
\,,\label{3decay-WW}\\[1ex]
\Gamma(h \to Z Z^*)
\!\!&=&\!\!
|C_{h WW}|^2 
\left[ 
\left( 6 - 12 s^2_W + \frac{152}{9} s^4_W \right) + 2 |C^V_{Zbb} \hat{v}_b|^2 + 2 |C^A_{Zbb} \hat{a}_b|^2
\right]
\frac{g^4 m_h}{2048\pi^3 c^4_W} F\left( \frac{M_Z}{m_h}\right)
\label{3decay-ZZ}\,, \\[1ex]
&&\hspace*{50ex} (\text{for $ M_Z \leq m_h \leq 2 M_Z$})
\,,\nonumber
\eeq
and the one-loop induced decay width $\Gamma(h \to \gamma \gamma/
gg)$ are given by \cite{Gunion:1989we}
\beq
\Gamma(h \to \gamma \gamma)
\!\!&=&\!\!
\frac{\alpha^2 g^2}{1024 \pi^3}\frac{m^3_h}{M^2_W}
\left| 
\begin{aligned}
&C_{hWW} A_1\left(\frac{4M^2_W}{m^2_h} \right)
+ \sum_{f} C_{hff} N_cQ^2_f A_{1/2} \left( \frac{4m^2_f}{m^2_h}\right)
\\
&+ \sum_{i = 1,2} C_{hH^\pm_i H^\pm_i} \frac{M^2_W}{M^2_{H^\pm_i}}A_0\left( \frac{4m^2_{H^\pm_i}}{m^2_h}\right)
\end{aligned}
\right|^2
\,,\label{1loop-gammagamma}
\\[1ex]
\Gamma(h \to gg)
\!\!&=&\!\!
\frac{\alpha^2_S g^2}{128 \pi^3}\frac{m^3_h}{M^2_W}
\left| 
\frac{1}{2}\sum_{t,b,T,B} C_{h ff}  A_{1/2} \left( \frac{4m^2_f}{m^2_h}\right)
\right|^2
\,,\label{1loop-gg}
\eeq
and $\Gamma(h \to Z\gamma )$ is given by \cite{Gunion:1989we,Abe:2012fb}
\beq
\Gamma(h \to Z \gamma)
\!\!&=&\!\!
\frac{\alpha^2 g^2}{512 \pi^3}\frac{m^3_h}{M^2_W} \left[ 1 - \frac{M^2_Z}{m^2_h}\right]^3
\left| 
\begin{aligned}
& C_{hWW} \frac{c_W}{s_W} A_1\left(\frac{4M^2_W}{m^2_h}, \frac{4M^2_W}{M^2_Z} \right) 
\\
& +  \sum_{f} C_{h ff} C^V_{Zff} \frac{2 N_c \hat{v}_f Q_f}{s_W c_W} A_{1/2} \left( \frac{4m^2_f}{m^2_h}, \frac{4m^2_f}{M^2_Z}\right) 
\\
&
+\frac{g}{c_W} \sum_{f\neq F}\left[ 
\begin{aligned}
&
\frac{4m_f}{m^2_F} \left( C^L_{hFf} C^L_{ZFf} +  C^R_{hFf} C^R_{ZFf}\right) 
\left( 3 + 4 \ln \frac{m_f}{m_F}\right)
\\
&
- \frac{4}{m_F} \left( C^L_{hfF} C^L_{ZFf} + C^R_{hfF} C^R_{ZFf}\right)
\end{aligned}
\right]
 \\
& +  \frac{c^2_W - s^2_W}{c_W s_W}\!\!\sum_{i=1,2}\!\! C_{hH^\pm_i H^\pm_i}  \frac{M^2_W}{M^2_{H^\pm_i}}A_0\left( \frac{4m^2_{H^\pm_i}}{m^2_\phi}, \frac{4m^2_{H^\pm_i}}{m^2_Z}\right)
\end{aligned}
\right|^2
\,,\label{1loop-Zgamma}
\eeq
where the third line in the right hand side of Eq.(\ref{1loop-Zgamma}) is satisfied for $m_F \gg m_f,m_h,M_Z$ \cite{Abe:2012fb}.
$F(x)$ is given by
\beq
F(x)
\!\!&=&\!\!
-|1 -x^2| \left(\frac{47}{2} x^2 - \frac{13}{2} + \frac{1}{x^2} \right)
+ 3(1-6 x^2 + 4x^4) |\ln x| 
+\frac{3(1-8x^2+20x^4)}{\sqrt{4x-1}}\arccos\left[ \frac{3x^2 - 1}{2x^3}\right]
\,,
\eeq
$A_{1,1/2,0}(x)$ are given by
\beq
A_1(x) 
\!\!&=&\!\! 
2 + 3x +3x (2-x)f(x)  \,, \\[1ex]
A_{1/2}(x) 
\!\!&=&\!\! 
2 x \left[ 1 + (1- x )f(x) \right]\,,
 \\[1ex]
A_{0}(x) 
\!\!&=&\!\! 
x [ x -f(x)]  \,, 
\eeq
and $A_{1,1/2,0}(x,y)$ are given by
\beq
A_1(x,y)
\!\!&=&\!\!
4(3-t^2_W) I_2(x,y) 
+ \left[\left( 1 + \frac{2}{x} \right) t^2_W - \left( 5 + \frac{2}{x}\right) \right] I_1(x,y)
\,,\\[1ex]
A_{1/2}(x,y)
\!\!&=&\!\!
 I_1(x,y) - I_2(x,y)
\,,\\[1ex]
A_{0}(x,y)
\!\!&=&\!\!
I_1(x,y) 
\,,\\[1ex]
I_1(x,y)
\!\!&=&\!\!
\frac{xy}{2(x-y)} + \frac{x^2y^2}{2(x-y)^2}\left[ f(x) - f(y)\right]
+ \frac{x^2y}{(x-y)^2} \left[ g(x) - g(y)\right]
\,,\\[1ex]
I_2(x,y)
\!\!&=&\!\!
-\frac{xy}{2(x-y)} \left[ f(x) - f(y)\right]
\,,
\eeq
where
\beq
f(x)
=
\left\{ \begin{array}{ll}
[\arcsin(1/\sqrt{x})]^2 & \text{for $x > 1$} \\[2ex]
-\dfrac{1}{4} \left[ \log \dfrac{1+\sqrt{1-x}}{1-\sqrt{1-x}} - i \pi \right]^2  & \text{for $x \leq 1$} \\[1ex]
\end{array} \right.
\quad , \quad
g(x)
=
\left\{ \begin{array}{ll}
\sqrt{x-1}\arcsin(1/\sqrt{x}) & \text{for $x > 1$} \\[2ex]
\dfrac{1}{2}\sqrt{1-x} \left[ \log \dfrac{1+\sqrt{1-x}}{1-\sqrt{1-x}} - i \pi \right] & \text{for $x \leq 1$} \\[1ex]
\end{array} \right.
\,.
\eeq


\end{document}